\documentclass[12pt,fleqn]{article}
\usepackage{graphicx,amsfonts,amsbsy,latexsym}
\def\eq#1\en{\begin{equation}#1\end{equation}}
\def\eqa#1\ena{\begin{eqnarray}#1\end{eqnarray}}
\def\junc#1\junc{} 
\newcommand*\nn{\nonumber \\}

\newlength{\vscaling} \newlength{\hscaling}
  \newcommand*\ti[5]{{\em #5}, {#1} {\bf #2}, #3 (#4)}

\newcommand*\np{Nucl. Phys.}

\def\Ax{\mathcal{A}_x} % noncommutative space-time

\newcommand*\Liecom[2]{\{ #1 \, , \, #2 \}_\mathrm{Lie}}

\newcommand*\scom[2]{[  #1 \, \stackrel{\star}{,} \, #2  ]}
\newcommand*\gcom[2]{[ #1 \, , \, #2  ]_\mathrm{G}}
\newcommand*\sncom[2]{[  #1 \, , \, #2 ]_\mathrm{S}}
\newcommand*\pcom[2]{\{ #1 \, , \, #2 \}}

\newcommand*\Hom{\mathrm{Hom}}

\newcommand*\dpo[1][]{\mathbf{d}_{\theta #1}} % Poisson
\newcommand*\ds[1][]{\mathbf{d}_{\star #1}}   % Star
\newcommand*\dst{\mathbf{\tilde d}_{\star}}

\newcommand*\ap[1][]{\mathbf{a}_{\theta #1}} % Poisson gauge potential
  \newcommand*\api[1][]{\mathbf{a}_{\vartheta #1}} % internal Poisson gauge potential
\newcommand*\as[1][]{\mathbf{a}_{\star #1}}   % Star gauge potential
\newcommand*\fp[1][]{\mathbf{f}_{\theta #1}} % Poisson field strength
\newcommand*\fs[1][]{\mathbf{f}_{\star #1}}   % Star field strength

\newcommand*\As[1][]{\hat A_{#1}}  % quantum
\newcommand*\Lq[1][]{\hat\lambda_{#1}}  % quantum
\newcommand*\LLq[1][]{\hat\Lambda_{#1}}  % quantum
\newcommand*\deltaq{\hat\delta}        % quantum transformation

\newcommand*\ha{\hat A}
\newcommand*\hl{\hat \lambda}

\newcommand*\btheta[1][]{\boldsymbol{\theta #1}}
\newcommand*\bomega[1][]{\boldsymbol{\omega #1}}
\newcommand*\wtheta{\tilde \theta}

\def\id{\mathrm{id}}

\def\pp{\partial}

\def\al{\alpha}

\def\BC{{\mathbb{C}}}

\newcommand*\EE{{\cal E}}
\def\AA{{\cal A}}
\def\FF{{\cal F}}

\def\MM{{\cal M}}

\def\DD{{\cal D}}
\def\OO{{\cal O}}
\def\CC{{\cal C}}

\begin{document}
\begin{titlepage}
%\rightline{MPI}
\rightline{LMU-TPW 00-20}
\vfill
\begin{center}
{\bf\LARGE
Nonabelian noncommutative gauge theory
via noncommutative extra dimensions}

\vfill

{{\bf Branislav Jur\v co, Peter Schupp and Julius Wess
}}

 \vskip 0.5 cm

Universit\"at M\"unchen,
Sektion Physik\\ Theresienstr.\ 37,
80333 M\"unchen, Germany

%\vskip 0.5 cm
%{\tt\footnotesize jurco\,,\,schupp\,,\,wess\,@\,theorie.physik.uni-muenchen.de}
\end{center}
\vfill
\begin{abstract} 
The concept of covariant coordinates on noncommutative
spaces leads directly to 
gauge theories with generalized noncommutative gauge
fields of the type that arises in string theory with background $B$-fields.
The theory is naturally expressed in terms of cochains in an appropriate
cohomology; we discuss how it fits into the framework of projective
modules. The equivalence of star products that arise
from the background field with and without fluctuations and
Kontsevich's formality theorem allow an explicitly construction of a map that 
relates ordinary gauge theory and noncommutative gauge theory 
(Seiberg-Witten map.) As application we show the exact equality of the
Dirac-Born-Infeld action
with $B$-field in the commutative setting and its semi-noncommutative
cousin in the intermediate picture. Using noncommutative
extra dimensions the construction is extended to
noncommutative nonabelian gauge theory for arbitrary gauge groups; an
explicit map between abelian and nonabelian gauge fields is given.
All constructions are also valid for non-constant $B$-field, Poisson
structure and metric.
\end{abstract}
\vfill
\hrule
\vskip 5pt
\noindent
{\footnotesize\it e-mail:
\parbox[t]{.8\textwidth}{jurco\,,\,schupp\,,\,wess\,@\,theorie.physik.uni-muenchen.de}}
\end{titlepage}\vskip.2cm

\newpage

\setcounter{page}{1}
%\thispagestyle{empty}
%%%%%%%%%%%%%%%%%%%%%%%%%%%%%%%%%%%%%%%%%%%%%%%%%%%%%%%%%%%%%%%%%%%%%%%
\tableofcontents
\pagebreak

\section{Introduction}
A natural approach to gauge theory on noncommutative spaces can be
based on the simple observation that multiplication of a field by a
(coordinate) function is not a covariant concept if that function does 
not commute with gauge transformations~\cite{MSSW}.  
This can be cured by adding appropriate 
noncommutative gauge potentials and thus introducing covariant functions 
in complete analogy to the covariant derivatives of ordinary gauge theory.%
\footnote{From the phase space point of view ordinary gauge
theory is in fact a special case of this construction with gauge
potentials  for only half of the `coordinates' (momenta).}
%Noncommutative geometry in general has enjoyed a recent surge 
This construction is of particular interest
because of its apparent relevance for the description of 
open strings in a
background $B$-field \cite{CLNY,CH,Scho}, where the D-brane world volume can be
interpreted as a noncommutative space whose fluctuations are governed by a noncommutative
version of Yang-Mills theory \cite{CDS,DH,MZ,Wati,SW}. 
It has been noticed (at least in the case of a constant $B$-field) that there
can be equivalent description of the effective theory both in terms of noncommutative gauge
theory and ordinary gauge
theory. From the physics perspective the two pictures are related by a choice of
regularization~\cite{SW,AD} which suggests a somewhat miraculous 
field redefinition that is usually
called Seiberg-Witten map~\cite{SW}. The inverse \mbox{$B$-field}, or more generally
the antisymmetric part of the inverse sum of $B$-field and metric, defines a
Poisson structure~$\theta$ whose quantization gives rise to the noncommutativity
on the D-brane world volume. Classically, the field strength $f$ describes
fluctuations of the $B$-field.
The Seiberg-Witten map expresses the noncommutative potential 
$\As$, 
noncommutative gauge 
parameter $\Lq$ and noncommutative field strength in terms of their 
classical counter parts $a$, $\lambda$, $f$ as formal power series in $\theta$, 
such that noncommutative gauge
transformations $\deltaq_{\Lq}$ are induced by ordinary gauge transformations
$\delta_\lambda$:
\eq
\As(a) + \deltaq_{\Lq}\As(a) = \As(a +\delta_\lambda a), \label{SWcond}
\en
where $\Lq$ is a function both of $a$ and $\lambda$. In previous
work~\cite{JS,JSW} we have focused on the rank one case and have explicitly 
constructed the maps $\As(a)$ and $\Lq(a,\lambda)$ to all orders in theta for
the general case of an arbitrary Poisson manifolds which are relevant
for the case of non-constant background fields.\footnote{This is of
course neither restricted to magnetic fields -- $B^{0i}$ need not be zero --
nor to even dimensional manifolds.} The corresponding
star products can be computed with Kontsevich's formula~\cite{Kontsevich};
this formula continues to make sense even for non-closed $B$-field
although the corresponding star product will no longer be associative
(see also~\cite{Cornalbana}) but the non-associativity is still under control
by the formality (\ref{control}).
For a noncommutative gauge theory 
the rank one case does already include some information
about the nonabelian case, since it is always possible to include
a matrix factor in the definition of the 
underlying noncommutative space. In this article we shall make this more
precise and will extend our previous results
to non-abelian gauge theories for \emph{any} Lie group.
(A brief description of the construction was  given in~\cite{JSW2}.)
The noncommutative gauge potential and field
strength in general take values in the universal enveloping algebra,
nevertheless thanks to the existence of the Seiberg-Witten map the theory 
can be consistently formulated in terms of only a finite number of fields;
this important observation has been discussed in 
~\cite{Jurco:2000ja}. A prerequisite for all this is an appropriate 
formulation of gauge theory on a more or less arbitrary noncommutative
space. (Here we are interested in the general case of an arbitrary
associative algebra of non-commuting variables,
important special examples with constant, linear and quadratic
commutation relations have been discussed in~\cite{MSSW}.) 
Particularly well-suited is the approach  based on the notion of 
covariant 
coordinates that we mentioned above
%
%\junc
%\footnote{This is a somewhat nonstandard formulation of
%noncommutative gauge theory that is not as intimately connected 
%with a differential calculus
%as it is in Connes approach. In particular examples, e.g. for the noncommutative
%torus, both formulations may be used and give the same results.}
%\junc
%
because it finds a natural interpretation in the framework of deformation 
quantization~\cite{BFFLS,Kontsevich,Sternheimer}. 
This is the natural setting since we are dealing with 
associative algebras and formal 
power series~-- it also allows rigorous statements by postponeing,
or rather circumventing difficult questions related to convergence and 
representation theory.
Deformation quantization for non-constant and possibly degenerate Poisson structures
goes far beyond the basic Weyl-Moyal product and the problem has only recently
found a general solution~\cite{Kontsevich}. To construct a Seiberg-Witten map 
we do in fact need the even
more general formality theorem of Kontsevich~\cite{Kontsevich}.
A link between Kontsevich quantization/formality  and quantum field theory
is given by the path integral approach~\cite{CattaneoFelder} which  relates the
graphs that determine the terms in the formality map to
Feynman diagrams. The relevant action~--
a Poisson sigma model~-- was originally studied
in~\cite{Schaller:1994es,Ikeda:1994fh};
see also \cite{Schaller:1994uj,Alekseev:1995py}.
Our discussion is entirely tree-level. Aspects of the quantization
of nonabelian noncommutative gauge theories have been discussed by several
authors~\cite{Bonora:2000ga,Armoni2,Armoni:2001br} (and references therein.)
Closest to the present discussion is the perturbative study of $\theta$-expanded
noncommutative gauge theories~\cite{Bichl:2001nf,Bichl2}.
%The paper is organized as follows: We review the concept of covariant
%coordinates on a noncommutative space which leads directly to noncommutative
%gauge theories with generalized noncommutative gauge fields.
%We then give an abstract mathematical formulation and show how all
%objects fit into the framework of projective modules
%which naturally take the place of fiber bundles in the noncommutative realm.

\subsection{Noncommutative gauge theory in string theory}
\label{sec:string}

Let us briefly recall how star products and
noncommutative gauge theory arise
in string theory~\cite{CH,Scho,SW}:
Consider an open string $\sigma$-model with background
term
\eq
S_B = \frac{1}{2i} \int_D B_{i\!j}\, \epsilon^{ab} \pp_a x^i \pp_b x^j ,
\en
where the integral is over the string world-sheet (disk) and
$B$ is constant, nondegenerate and $dB = 0$. The correlation functions
on the boundary of the disc in the decoupling limit ($g \rightarrow 0$,
$\alpha' \rightarrow 0$) are
\eq
\left\langle f_1(x(\tau_1)) \cdot\ldots\cdot f_n(x(\tau_n))\right\rangle_B
= \int dx \, f_1 \star \ldots
\star f_n , \quad (\tau_1 < \ldots < \tau_n) \label{corr}
\en
with the Weyl-Moyal star product
\eq
(f \star g ) (x)  = \left. e^{\frac{i\hbar}{2}\theta^{i\!j} \pp_i \pp_j'} f(x) g(x') \right|_{x'
\rightarrow x} ,
\en
which is the deformation quantization of the Poisson structure $\theta =
B^{-1}$.
More generally a star product is an associative, $[[\hbar]]$-bilinear product
\eq
f \star g = f g + \sum_{n=1}^\infty (i\hbar)^n
\underbrace{B_n(f,g)}_{\mathrm{bilinear}} ,
\en
which is the deformation of a Poisson structure $\theta$:
\eq
\scom{f}{g} = i \hbar \pcom{f}{g} + \OO(\hbar^2), \quad \pcom{f}{g} =
\theta^{i\!j}(x) \pp_i f \, \pp_j g .
\en
We now perturb the constant $B$ field by adding a gauge potential $a_i(x)$:
$B \rightarrow B + da$, $S_B \rightarrow S_B + S_a$, with
\eq
S_a = -i \int_{\partial D} d\tau a_i(x(\tau))\partial_\tau x^i(\tau).
\en
Classically we have the naive gauge invariance
\eq
\delta a_i = \partial_i \lambda, \label{naive}
\en
but in the quantum theory this depends on the choice of regularization.
For Pauli-Villars (\ref{naive}) remains a symmetry but
if one expands $\exp S_a$ and employes a point-splitting regularization
then the functional integral is invariant under noncommutative gauge
transformations\footnote{In this form this formula is 
only valid for the Moyal-Weyl
star product.}
\eq
\hat\delta \ha_i = \partial_i \hl + i \hl \star \ha_i - i \ha_i \star \hl.
\label{nctrans}
\en
Since a sensible quantum theory should be independent of the choice of regularization
there should be field redefinitions 
$\ha(a)$, $\hl(a,\lambda)$ (Seiberg-Witten map) that relate (\ref{naive}) and
(\ref{nctrans}):
\eq
\ha(a) +\hat\delta_{\hl} \ha(a) = \ha(a+\delta_\lambda a). \label{swcond1}
\en
It is instructive to study the effect of the extra factor $\exp S_a$ in
the correlation function (\ref{corr})
in more detail: It effectively shifts the
coordinates\footnote{Notation: $\DD$ should not be confused with a 
covariant
derivative (but it is related).}
\eq
x^i \rightarrow x^i + \theta^{ij}\ha_j =: \DD x^i . \label{covco1}
\en
More generally, for a function $f$,
\eq
f \rightarrow f + f_A =: \DD f. \label{covfu}
\en
The mapping $\AA: \, f \mapsto f_A$ plays the 
role of a generalized gauge potential; it maps a function
to a new function that depends on the gauge potential.
The shifted coordinates and functions are covariant under noncommutative
gauge transformations:
\eq
\hat\delta (\DD x^i) = i\scom{\hl}{\DD x^i}, \qquad
\hat\delta (\DD f) = i\scom{\hl}{\DD f}.
\en
The first expression implies (\ref{nctrans}) (for $\theta$ constant and
nondegenerate).

The covariant coordinates (\ref{covco1}) are the
background independent operators of \cite{SW,Seiberg}; 
they and the covariant functions (\ref{covfu}) can also
be introduced abstractly in the general case of an
arbitrary noncommutative space as we shall discuss
in the next section.

\section{Gauge theory on noncommutative spaces}

\subsection{Covariant functions, covariant coordinates}

Take a more or less arbitrary noncommutative space, i.e.\ an
associative unital algebra $\Ax$ of noncommuting 
variables with multiplication~$\star$
and consider (matter) fields $\psi$ on this space.
The fields  can be taken to be elements of $\Ax$, or, more generally, 
a left module of it.
The notion of a gauge transformation is introduced 
as usual\footnote{We shall
often use the
infinitesimal version $\delta \psi = i\lambda\star\psi$ 
of~(\ref{gaugetrafo}) -- this is
purely for notational clarity. Other transformations, like, e.g.,
$\psi \mapsto \psi\star\Lambda$ or $\psi \mapsto \Lambda\star\psi\star\Lambda^{-1}$
can also be considered.}
\eq
\psi \mapsto \Lambda \star \psi , \label{gaugetrafo}
\en
where $\Lambda$ is an invertible element of $\Ax$.
In analogy to commutative geometry where a manifold can be described by the
commutative space of functions over it, we shall refer to the elements of $\Ax$
also as functions. Later we shall focus on the case where the noncommutative
multiplication is a star product; the elements of $\Ax$ are then in fact
ordinary functions in the usual sense of the word.
The left-multiplication  of a field with a function $f \in \Ax$ does in general
not result in a covariant object because of the noncommutativity of $\Ax$:
\eq
f\star\psi \mapsto f\star\Lambda\star\psi \neq \Lambda\star (f \star\psi) .
\label{noncov}
\en
(As in ordinary gauge theory the gauge transformation only acts on the
fields, i.e.\ on the elements of the left-module of $\Ax$ and not on 
the elements of $\Ax$ itself.) To cure (\ref{noncov}) we introduce covariant functions
\eq
\DD f = f + f_A ,
\label{covf}
\en
that transform under gauge transformations by conjugation
\eq
\DD f \mapsto \Lambda\star \DD f \star \Lambda^{-1} , \label{covft}
\en
by adding `gauge potentials' $f_A$ with appropriate transformation
property\footnote{Notation: $\scom{a}{b} \equiv a\star b - b \star a \equiv
[a,b]_\star$.}
\eq
f_A \mapsto \Lambda\star \scom{f}{\Lambda^{-1}} + \Lambda\star f_A \star
\Lambda^{-1} . \label{potentials}
\en
Further covariant objects can be constructed from covariant functions; 
the `2-tensor'
\eq
\FF(f,g) = \scom{\DD f}{\DD g} - \DD(\scom{f}{g}) , \label{tensor}
\en
for instance plays the role of covariant noncommutative field strength. 

\subsubsection{Canonical structure (constant $\mathbf \theta$) 
and noncommutative Yang-Mills}

Before we continue let
us illustrate all this in the particular simple case of an algebra $\Ax$ generated
by `coordinates' $x^i$ with canonical commutation relations
\eq
\scom{x^i}{x^j} = i\theta^{ij} , \quad \theta^{ij} \; \in \; \BC.
\label{canonical}
\en
This algebra arises in the decoupling limit of open strings in 
the presence of a \emph{constant} $B$-field. It
can be viewed as the quantization of a Poisson structure with
Poisson tensor $\theta^{ij}$ and the multiplication $\star$ is then the
Weyl-Moyal star product
\eq
f\star g = f e^{\frac{i}{2}\theta^{ij}
\stackrel{\leftarrow}{\partial_i} \otimes \stackrel{\rightarrow}{\partial_j}} g.
\en
(This formula holds only in the present example, where $\theta^{ij}$ is
constant and we shall also assume that it is non-degenerate. In the rest
of the paper we drop both restrictions.) Let us focus on the coordinate functions
$x^i$. The corresponding covariant coordinates are
\eq
\DD x^i = x^i + x_A^i = x^i + \theta^{ij} \hat A_j, \label{covco}
\en
where we have used $\theta$ to lower the index on $\hat A_j$. 
Using (\ref{canonical}), we see that 
the transformation~(\ref{potentials}) of the 
noncommutative gauge potential $\hat A_j$ is 
\eq
\hat A_j \mapsto i\Lambda \star \pp_j(\Lambda^{-1}) + \Lambda \star \hat A_j
\star \Lambda^{-1},  \label{ncgt}
\en
or, infinitesimally
\begin{equation}
  \delta \hat A_j = \pp_j\lambda + i[\lambda,\hat A_j]_\star .
\end{equation}
The noncommutative field strength
\eq
\hat F_{kl} = \pp_k \hat A_l - \pp_l \hat A_k - i [\hat A_k,\hat A_l]_\star 
\label{ncfs}
\en
transforms covariantly 
\eq
\hat F_{kl} \mapsto \Lambda \star \hat F_{kl} \star \Lambda^{-1} .
\label{ncft}
\en
We have again used $\theta$ to lower indices to get
(\ref{ncfs}) from the definition (\ref{tensor})
\eq
i\hat F_{kl} \theta^{ik}\theta^{jl} \equiv
\FF(x^i,x^j) = [x_A^i,x^j]_\star + [x^i,x_A^j]_\star + [x_A^i,x_A^j]_\star . \label{covF}
\en
Note, that we should in general be more careful when using $\theta$ to 
lower indices as in (\ref{covco}) or (\ref{covF}) because this may spoil 
the covariance when $\theta$ is not constant as it was in this particular example. 
Relations (\ref{ncgt}), (\ref{ncfs}) and (\ref{ncft}) define what is usually called
Noncommutative Yang-Mills theory (NCYM) in the narrow sense: ordinary Yang-Mills
with all matrix products replaced by star products. This simple rule, however, only
really  works well for the Moyal-Weyl product, i.e.\ constant $\theta$.
In the general case it is wise to stick with the manifestly covariant and
coordinate-independent\footnote{We would like to thank Anton Alekseev for stressing 
the importance of this point.} objects defined in 
(\ref{covf}) and (\ref{tensor}). 
The fundamental objects  are really
the mappings (differential operators) $\DD$ and $\FF$ in these equations. 
The transformation of
$\AA = \DD - \id: f \mapsto f_A$
under gauge transformations is exactly
so that (\ref{covf})
transforms by conjugation.
The mappings $\AA \in \Hom(\Ax,\Ax)$ and $\FF \in \Hom(\Ax\wedge\Ax,\Ax)$ play the role
of generalized noncommutative gauge potential and noncommutative field strength.
There are several reasons, why one needs $\AA$ and $\DD$ and not just
$A^i \equiv \AA(x^i)$ (or $\hat A_i$, for $\theta$ constant):
If we perform a general coordinate transformation
$x^i \mapsto {x^i}'(x^j)$ and
transform $A^i$ (or $\hat A_i$) naively as its index structure suggests,
then we would obtain objects that are no longer covariant under
noncommutative gauge transformations. The correct transformation,
$\AA(x^i) \mapsto \AA'({x^i}')$, is more complicated and will be
discussed in section~\ref{sec:coordtransform}. 
Furthermore we may be interested
in covariant versions of scalar fields $\phi(x)$. These are given
by the corresponding covariant function $\DD(\phi(x))$.

In the next section we will propose an abstract definition of
the type of noncommutative gauge theory that is of present interest. Then we shall
proceed to give an interpretation in the framework of 
deformation quantization  
and will construct a particular important class of these operators.

\subsection{Abstract formulation of noncommutative gauge theory}

Finite projective modules take the place of fiber bundles in the noncommutative
realm \cite{Connes}. This is also the case here, as we shall explain below,
but may not have been apparent 
since we have been working with component
fields as is customary in the physics literature.
We have argued in the previous section that $\AA \in C^1$, $\FF \in C^2$ with
$C^p = \Hom(\Ax^{\wedge p},\Ax)$, $C^0 \equiv \Ax$.
These $p$-cochains take the place of forms
on a noncommutative space $\Ax$, which for now is still an arbitrary associative 
algebra over a field $k$ with
multiplication~$\star$.
It is actually more convenient to start with the Hochschild complex
of $\Ax$, 
$H^p(\Ax,\Ax) = \Hom_k(\Ax^{\otimes p},\Ax)$,
with values in $\Ax$ considered as a left module of $\Ax$. 
(The formulas for $C^p$ can then be obtained by antisymmetrization.)
We have a coboundary operator
$\ds: H^p \rightarrow H^{p+1}$, $\ds^2 = 0$, $\ds 1 = 0$,
\eq
\ds \mathcal{C} = -\gcom{\mathcal{C}}{\star\,}, \label{dq}
\en
where $\gcom{}{}$ is the Gerstenhaber bracket (\ref{gerstenhaber}),
\eqa
\lefteqn{(\ds \mathcal{C})(f_1,\ldots,f_{p+1})}&& \nn
 && =  f_1 \star \mathcal{C}(f_2,\ldots,f_{p+1}) 
   -\mathcal{C}(f_1\star f_2, f_3, \ldots , f_{p+1})
   + \mathcal{C}(f_1, f_2\star f_3, \ldots, f_{p+1}) \nn
 && \quad \mp \cdots 
   +(-)^{p} \mathcal{C}(f_1, f_2, \ldots,f_p\star f_{p+1}) 
   +(-)^{p+1} \mathcal{C}(f_1,\ldots,f_p) \star f_{p+1} , \label{hochschilddc}
\ena
and we have a cup product
$\star: H^{p_1} \otimes H^{p_2} \rightarrow H^{p_1 + p_2}$,
\eq
\left(\mathcal{C}_1\star\mathcal{C}_2\right)\left(f_1, %\ldots,f_{p_1},f_{p_1+1},
\ldots,f_{p_1+p_2}\right)
= \mathcal{C}_1(  f_1,\ldots,f_{p_1} ) \star \mathcal{C}_2( f_{p_1+1},
    \ldots,f_{p_1+p_2}  ) .
\en
For a function $\lambda\in H^0 \equiv \Ax$ the coboundary operator is defined
as 
\eq
(\ds\lambda)(f) = f\star\lambda - \lambda\star f \label{hochschilddonf}
\en 
and the cup
product reduces to the multiplication $\star$ of $\Ax$ in the obvious way.
For this reason and since there seems to be little chance of confusion we have
used the same symbol $\star$ for the cup product and the
multiplication.
Let us apply the Hochschild formalism to the gauge transformation dependent
map $\DD \in H^1$ that we introduced in the definition of 
covariant functions (\ref{covf}) in the previous section. In view of the way that
$A^i \equiv \theta^{ij} \hat A_j$ appeared in the definition of covariant
coordinates~(\ref{covco}) we define an abstract noncommutative
gauge potential $\AA \in H^1$
\eq
\AA \equiv \DD - \id. \label{nca}
\en
Applying $\AA$ to coordinate functions in the setting of (\ref{covco}) 
we indeed 
recover $A^i$: 
\[
\AA(x^i) = \DD(x^i) - x^i =  A^i .
\]
Let us compute the behavior of $\AA$ under a gauge transformation. Using
(\ref{covft}) 
and the definitions of $\ds$ and
the cup product we find 
%the following line temporarily changes \cup -> \star. 
{\renewcommand\cup{\star}
\eq
\AA \mapsto \Lambda\star\ds\Lambda^{-1} + \Lambda\star\AA\star\Lambda^{-1},
\en
which gives (\ref{potentials}) when evaluated on a function.
The corresponding infinitesimal version is
\eq \label{deltaA}
\delta\AA =  i(-\ds\lambda + \lambda\cup\AA - \AA\cup\lambda).
\en
Next we introduce the ``Hochschild'' field strength $\FF_H \in H^2$ 
\eq  \label{FHS}
\FF_H \equiv \ds \AA + \AA\cup\AA 
\en
and compute
\eq
\FF_H(f,g) = \DD f \star \DD g - \DD(f\star g) \label{symF} 
\en
and find the Bianchi identity
\eq \label{dF}
\ds \FF_H + \AA\cup\FF_H - \FF_H\cup\AA = 0.
\en
Evaluated on three functions $f$, $g$, $h$ the latter reads
\eq
\DD((f\star g)\star h) - \DD(f\star(g\star h))
+(\DD f\star \DD g) \star \DD h -
\DD f \star (\DD g \star \DD h),
\en
which is zero by associativity of $\Ax$.
$\FF_H$ transforms covariantly under gauge transformations
\eq
\FF_H \mapsto \Lambda\star\FF_H\star\Lambda^{-1};
\en
infinitesimally
\eq \label{deltaF}
\delta\FF_H = i (\lambda\cup\FF_H - \FF_H\cup\lambda).
\en
}%
When we compare
(\ref{tensor}) and (\ref{symF}), we see (as expected) that our noncommutative field 
strength $\FF$ of the 
previous section  is an antisymmetric version of $\FF_H$. 
We can obtain $\FF$ directly by taking the antisymmetrized version of
Hochschild, where one considers $\Ax$ as a Lie algebra with
bracket $\scom{a}{b} = a\star b - b\star a$; this is the Chevalley
cohomology of $\Ax$ with values in $\Ax$: $C^p = \Hom_k(\Ax^{\wedge p},\Ax)$.
We find the relevant formulas in this setting by replacing $H^p$ with $C^p$
(whose elements are 
antisymmetric), and
by using corresponding antisymmetrized
formulas for the coboundary operator $\ds$ and the cup product
which we then denote by $\wedge$. The action of
$\mbox{Lie} \Ax$ on the module $\Ax$ is given by $\star$-multiplication as before.
We now see that equation (\ref{tensor}),
\eq
\FF(f,g) = \scom{\DD f}{\DD g} - \DD(\scom{f}{g}), \label{antisymF} 
\en
can be written
\eq
\FF \equiv \ds \AA + \AA\wedge\AA. \label{asymF}
\en
The remaining equations also do not change in form (as compared to the Hochschild case): (\ref{asymF}) implies
\eq
\ds \FF + \AA\wedge\FF - \FF\wedge\AA = 0 \label{dasymF}
\en
and the behavior under (infinitesimal )gauge transformations is
\eqa
\delta\AA &= & i(-\ds\lambda + \lambda\wedge\AA - \AA\wedge\lambda) , \label{deltaaA} \\
\delta\FF &= & i\left( \lambda\wedge\FF - \FF\wedge\lambda\right). \label{deltaaF}
\ena
%There are of course also corresponding finite versions.
Equations (\ref{asymF}), (\ref{dasymF}), (\ref{deltaaA}) and (\ref{deltaaF}) 
are reminiscent of
the corresponding equations of ordinary (nonabelian) gauge theory. The correspondence
is given by the following dictionary: One-forms become linear operators on $\Ax$
which
take one function as argument and yield a new function,
two-forms become bilinear operators on $\Ax$ which take two functions as arguments
and return one new function, and the Lie bracket is replaced
by the antisymmetrized cup product $\wedge$.\footnote{More educated:
$n$-forms become $n$-cochains. An even closer match with the usual
physics conventions is achieved by multiplying our $\ds$ and $\FF$ by $i$
(section~\ref{ordinary} and later: multiply by $i\hbar$).}
As in ordinary gauge theory, it may
not be possible to use one globally defined gauge potential $\AA$; we may need
to introduce several $\DD$ and corresponding gauge potentials $\AA$ for
functions defined on different ``patches''. We shall come back to this later.
 
\noindent {\small \emph{Remark:}
One reason for going through the slightly more general Hochschild 
construction first is that
the symmetric part of $\FF_H$ may also contain interesting information
as we will see in section~\ref{metric}.
For invertible $\DD$ there is still another interesting object:
\eq
\widetilde\FF \equiv \DD^{-1} \circ \FF
\en
measures noncommutativity:
\eq
\widetilde\FF(f,g) = [f\stackrel{\star'}{,} g] - [f\stackrel{\star}{,} g], \label{ftilde}
\en
where the (associative) product $\star'$ is defined by
$f\star' g = \DD^{-1}\Big(\DD f \star \DD g\Big)$.
This ``field strength'' satisfies the Cartan-Maurer equation
\eq
\ds\widetilde\FF = \gcom{\widetilde\FF}{\widetilde\FF}.
\en}

\subsubsection{Projective modules}

We shall now discuss how our formulae fit into the framework of finite
projective modules: 
The calculus of $p$-cochains in $C^p$ with the
coboundary operator $\ds$ uses only the algebraic structure of $\Ax$; it is
related to the standard universal calculus and one can obtain
other calculi by projection.
Consider a (finite) projective right $\Ax$-module $\EE$.
We introduce a connection on $\EE$ as a linear map
$\nabla : \EE \otimes_{\Ax} \!C^p \rightarrow \EE \otimes_{\Ax} C^{p+1}$
for $p \in \mathbb{N}_0$
which satisfies the Leibniz rule
\eq
\nabla(\eta \psi)
 =  (\tilde\nabla \eta) \psi
+ (-)^p \eta\,\dst\psi \nonumber
\en
for all $\eta\in\EE \otimes_{\Ax} \!C^p$, $\psi \in C^r$, and where
$\tilde\nabla\eta = \nabla\eta - (-)^p \eta\,\dst 1$,
\eq
\dst(a\wedge\psi) = (\ds a)\wedge\psi + (-)^q a \wedge(\dst\psi)
\en
for all $a \in C^q$, and $\dst 1$ is the identity operator on $\Ax$.
(The transformation of matter
fields $\deltaq \psi = i \hat\lambda\star\psi$ leads to a
slight complication here; for fields
that transform in the adjoint (by star-commutator) we would only need $\tilde\nabla$,
$\ds$ and not $\nabla$ and $\dst$.) 
%\eq
%(\nabla\eta)(f_1\wedge\ldots\wedge f_{p+1})
%= (\tilde\nabla\eta)(f_1\wedge\ldots\wedge f_{p+1}) \pm
%\en
%\eqa
%\nabla(\eta a \psi)
%& = & (\tilde\nabla \eta) (a\!\star\psi)
%+ (-)^p \eta\,\dst(a\!\star\psi) \nonumber \\
%& = & (\tilde\nabla \eta) (a\!\star\psi) + (-)^p \eta (\ds a) \psi + (-)^{p+q} \eta a
%(\dst\psi)
%\ena
%for all $\eta\in\EE \otimes_{\Ax} \!C^p$, $a \in C^q$, $\psi \in C^r$, and where
%$\tilde\nabla\eta = \nabla\eta - (-)^p \eta\dst 1$ and
%\eqa
%\dst \psi(f_1,\ldots,f_{r+1}) & = & f_1\star\psi(f_2,\ldots,f_{r+1})\nonumber \\
%&&+ \sum
%\psi_{(k)}(f_1,\ldots,f_{r+1})
%(-)^k\psi(f_1,\ldots,f_k\star
%f_{k+1},\ldots,f_{r+1}).
%\ena

Let $(\eta_a)$ be a  generating family for $\EE$;
any $\xi \in \EE$ can then be written as $\xi = \sum \eta_a \psi^a$ with $\psi^a \in \Ax$
(with only a finite number of terms different from zero). For a free module
the $\psi^a$ are unique, but we shall not assume that. Let the generalized
gauge potential be defined by the action of $\tilde\nabla$ on the elements
of the generating family: $\tilde\nabla \eta_a = \eta_b \AA^b_a$. In the
following we shall suppress indices and simply write $\xi = \eta.\psi$,
$\tilde\nabla \eta = \eta.\AA$ etc. We compute
\eq
\nabla\xi = \nabla(\eta.\psi) = \eta.(\AA\wedge\psi + \dst \psi)
= \eta.(\DD\wedge\psi).
\en
Evaluated on a function $f\in\Ax$ the component $\DD\wedge\psi$
yields a covariant function
times the matter field, $(\DD\wedge\psi) (f) = (\DD f)\star\psi$,
so in this framework covariant functions are related to the covariant
``derivative''
$\dst + \AA$:
\eq
[(\dst + \AA)\psi](f) = (f + \AA(f))\star\psi .
\en
The square of the connection gives
\eq
\nabla^2 \xi = \eta.(\AA\wedge\AA + \ds\AA).\psi = \eta.\FF.\psi
\en
with the field strength
\eq
\FF =  \ds\AA + \AA \wedge \AA.
\en

\section{Ordinary versus noncommutative gauge theory}
\label{ordinary}

We are particularly interested in the case where the algebra of our noncommutative
space $\Ax$ is given by a star product (via a quantization map).
A star product on a smooth $C^\infty$-Manifold $\MM$ is an associative
$\BC[[\hbar]]$-bilinear product
\eq
f \star g = f g + \sum_{n=1}^\infty \left(\frac{i\hbar}{2}\right)^n B_n(f,g) ,
\quad f,g \in C^\infty(\MM) ,
\en
where $B_n$ are bilinear operators and $\hbar$ is the formal deformation 
parameter; it is a deformation quantization of
the Poisson structure
\eq
\{ f, g\} \equiv \theta^{ij}(x) \pp_i f \, \pp_j g = B_1(f,g) - B_1(g,f) .
\en
Equivalent star products $\tilde\star$ can be constructed with the help
of invertible operators $D$
\eq
D(f \,\tilde\star\, g) = D f \star D g , \qquad D f \equiv f + \sum_{n=1}^\infty \hbar^n
D_n(f), \label{equiv}
\en
where $D_n$ are linear operators. This operation clearly does not spoil
associativity. There are also inner automorphisms for each invertible 
element~$\Lambda$
and their infinitesimal version, inner derivations,
\eq
f \mapsto \Lambda \star f \star \Lambda^{-1} , 
\qquad \delta f = \scom{i \lambda}{f}; \label{iauto}
\en
these operations do not change the star product.

The striking similarity between equations (\ref{covf}), (\ref{covft}) and equations
(\ref{equiv}), (\ref{iauto}) suggests the following interpretation of noncommutative gauge
theory in the star product formalism:
The covariance maps $\DD = \id + \AA$ are gauge equivalence maps $D$ for the
underlying star product,
combined with a change of coordinates $\rho^*$
\eq
\DD = D \circ \rho^*, \qquad \DD (f\star' g) = \DD f \star \DD g ;
\en
gauge transformations are inner automorphisms of the star
products.

\subsection{Motivation from string theory} 

A Poisson tensor $\theta$ enters the discussion of Seiberg and 
Witten~\cite{SW} via a background 
$B$-field in the open string picture. In this setting
\eq
\theta^{ij} = 2\pi\alpha'\left(\frac{1}{g+2\pi\alpha' B}\right)^{ij}_A 
\en
appears in the propagator at boundary points of the string world sheet.
($g$ is the closed string metric and $A$ denotes the antisymmetric
part of a matrix.)
The 2-form
$\omega \equiv \frac{1}{2} B_{ij} dx^i\wedge dx^j$ is a symplectic form, provided
$B$ is nondegenerate and
$d\omega = 0$, which is e.g.\ obviously the case if $B$ is constant (but we
shall not require it to be constant.)
In the zero slope limit (or in the intermediate picture with $\Phi = -B$~\cite{SW},
see section~\ref{DBI})
\eq
\theta^{ij} = (B^{-1})^{ij}
\en
defines then a Poisson structure. It has been discussed by several authors how
the Moyal-Weyl star product enters the picture as a quantization of this
Poisson structure in the constant case~\cite{CH,Scho,SW}. 
A direct approach that is most suitable for our
purposes and that also works for non-constant $\theta$ is given by 
Cattaneo and Felder~\cite{CattaneoFelder}
in their QFT realization of Kontsevich's star product.

In section~\ref{sec:string} we have discussed what happens if we add fluctuations $f$ (with
$df = 0$, i.e. locally $f = da$) to the background $B$ field: The action
is then naively invariant under ordinary gauge transformations $\delta a = d\lambda$,
but the invariance of the quantum theory depends on the choice of
regularization. A point splitting prescription \cite{SW,AD} leads in fact to
a noncommutative gauge invariance. Since in a consistent quantum theory the choice of 
regularization should not matter, Seiberg and Witten argued that there should
exist maps that relate ordinary and noncommutative gauge theory such that
(\ref{SWcond}) holds.
A more abstract argument that leads to the same conclusion, but gives the 
Seiberg-Witten maps more directly and also works for
non-constant $\theta$
can be based on a quantum version of Moser's lemma \cite{JS,JSW}.
Here is briefly the idea, in the next section we will review the details:
The addition of $f = da$ to $B$  defines
a new Poisson structure
\eq
\theta = \frac{1}{B} \quad\rightarrow\quad \theta' = \frac{1}{B + f}
\label{BplusF}
\en
which, according to Moser's lemma, is related to the original one by a change of
coordinates given by a flow $\rho^*_a$ that depends on the gauge potential $a$.
After quantization $\theta$ and $\theta'$ give rise to equivalent
star products $\star$ and $\star'$.
The equivalence map $D_a$, the full quantum flow
$\DD_a = D_a \circ \rho^*_a$
and the noncommutative gauge potential 
\eq
%\AA(a) = \DD_a - \id
\AA_a = \DD_a - \id
\en
are also functions of $a$.
An additional infinitesimal gauge transformation 
\eq
\delta a = d\lambda
\en
does not change the Poisson structure (since $\delta f = 0$),
but it still induces an infinitesimal canonical transformation.
After quantization that transformation becomes an inner derivation of the
star product $\star$ and thus a noncommutative gauge
transformation
$\delta_{\hat\lambda}$  with $\hat\lambda = \hat\lambda(\lambda,a)$,
such that
\eq
%\AA(a + d\lambda) = \AA(a) + \hat\delta_{\hat\lambda}\AA(a).
\AA_{a + d\lambda} = \AA_a + \delta_{\hat\lambda}\AA_a . \label{swcond}
\en

\subsection{Quantum version of Moser's lemma: Seiberg-Witten map}

Consider an abelian gauge theory on a manifold that also carries
a Poisson structure~$\theta$. The gauge potential, field strength
and infinitesimal gauge transformations are
\eq
a = a_i dx^i, \qquad f= \frac{1}{2}f_{ij}\,dx^i\wedge dx^j = da,
\qquad f_{ij} = \pp_i a_j - \pp_j a_i, \qquad \delta_\lambda a = d\lambda.
\en
We will first construct a semiclassical version of the Seiberg-Witten map,
where all star commutators are replaced by Poisson brackets. The construction
is essentially a formal generalization of Moser's lemma to Poisson manifolds.

\subsubsection{Semi-classical construction}
\label{sec:semiclass}

Let us consider the  nilpotent coboundary operator of
the Poisson cohomology (see \cite{Weinstein}) -- the semiclassical limit
of (\ref{dq}) -- 
\eq
\dpo = -\sncom{\,\cdot\,}{\theta},
\en
where $\sncom{}{}$ is the Schouten-Nijenhuis bracket (\ref{sncom})
and $\theta = \frac{1}{2} \theta^{ij} \partial_i\wedge\partial_j$ is
the Poisson bivector.
Acting with $\dpo$ on a function $f$ gives the Hamiltonian vector field corresponding
to $f$
\eq
\dpo f =  \{\,\cdot\,,f\} = \theta^{ij}(\pp_j f)\pp_i. \label{hamvec}
\en
It is natural to introduce a vector field 
\eq
\ap = a_i \dpo x^i = \theta^{ji} a_i \pp_j \label{vf}
\en
corresponding to the abelian gauge
potential $a$ and a bivector field
\eq
\fp = \dpo\ap = - \frac{1}{2}\theta^{ik} f_{kl} \theta^{lj}\, \pp_i\wedge\pp_j
\label{fs}
\en
corresponding to the abelian field strength $f = da$.
We have $\dpo\fp = 0$,
due to $\dpo^2 \propto \sncom{\theta}{\theta} = 0$
(Jacobi identity).

We are now ready to perturb the Poisson structure $\theta$ by introducing
a one-parameter deformation $\theta_t$ with $t \in [0,1]$:%
\footnote{In this
notation the equations resemble those of Moser's original lemma, which deals with
the symplectic 2-form $\omega$, the inverse of $\theta$ (provided it exists).
There, e.g., $\pp_t \omega_t = f$ for $\omega_t = \omega + t f$.}
\eq
\pp_t \theta_t = \fp[_t]  \label{evolution}
\en
with initial condition $\theta_0 =
\theta$. In local coordinates:
\eq
\pp_t \theta_t^{ij} = -(\theta_t f \theta_t)^{ij} ,\qquad \theta_0^{ij} = \theta^{ij},
\label{partialtf}
\en
with formal solution given by the geometric series
\eq
\theta_t = \theta - t \theta f \theta + t^2 \theta f \theta f \theta
- t^3 \theta f \theta f \theta f \theta \pm \cdots
= \theta \frac{1}{1 + t f \theta},
\en
if $f$ is not explicitly $\theta$-dependent.
(The differential equations (\ref{evolution}), (\ref{partialtf}) and the rest
of the construction
do make sense even if $f$ or $a$ are
$\theta$-dependent).
$\theta_t$ is a Poisson tensor for all $t$ because
$\sncom{\theta_t}{\theta_t}= 0$ at $t=0$ and
\[
\pp_t\sncom{\theta_t}{\theta_t}
%= \sncom{\fp[_t]}{\theta_t} + \sncom{\theta_t}{\fp[_t]}
= -2\dpo[_t]\fp[_t] \propto \sncom{\theta_t}{\theta_t} .
\]
The evolution (\ref{evolution}) of $\theta_t$ is generated by the vector field $\ap$:
\eq
\pp_t\theta_t = \dpo[_t]\ap[_t] = -\sncom{\ap[_t]}{\theta_t}.
\en
This Lie derivative can be integrated to a flow (see
appendix~\ref{s:t-evolution})
\eq
\rho^*_a = \left.\exp(\ap[_t] + \pp_t)\exp(-\pp_t)\right|_{t=0} \label{flow}
\en
that relates the Poisson structures $\theta' = \theta_1$ and $\theta = \theta_0$.
In analogy to (\ref{nca}) we define a semi-classical (semi-noncommutative)
generalized gauge potential
\eq
%\Ap(a) = \rho^*_a - \id.
A_a = \rho^*_a - \id. \label{as}
\en
Under an infinitesimal gauge transformation $a \mapsto a + d\lambda$ the
vector field (\ref{vf}) changes by a Hamiltonian vector field $\dpo\lambda
= \theta^{ij} (\pp_j\lambda)\pp_i$:
\eq
\ap \mapsto \ap + \dpo\lambda. \label{astrafo}
\en
Let us compute the effect of this gauge transformation on the flow (\ref{flow}).
%For this it is helpful to first evaluate
%\[
%\left[\pp_t + \ap , \dpo\lambda\right]
%= (\pp_t\dpo)\lambda + \dpo\ap(\lambda) - (\dpo\ap)\lambda =
%\dpo\ap(\lambda) =  \theta^{ij}\pp_j(\theta^{kl}a_l\pp_k\lambda)\pp_i,
%\]
%which is again a Hamiltonian vector field;
%the first and last terms canceled due to (\ref{fs}) and (\ref{evolution}).
After some computation (see appendix~\ref{lambdatilde})
we find (infinitesimally: to first order in
$\lambda$)
\eq
\rho^*_{a+d\lambda}
=  (\id + \dpo\tilde\lambda)\circ\rho^*_a,
\quad\mbox{i.e.,}\quad
\rho^*_{a+d\lambda}(f)= \rho^*_a(f) + \{\rho^*_a(f),\tilde\lambda\}
\label{ctrafo}
\en
and
\eq
A_{a+d\lambda} = A_a + \, \dpo\tilde\lambda + \{A_a,\tilde\lambda\},
\label{cswcond}
\en
%where $(\dpo\tilde\lambda + \{A_a,\tilde\lambda\})(f)
%= \{f,\tilde\lambda\} + \{A_a(f),\tilde\lambda\}$
with
\eq
%\tilde\lambda =
\tilde\lambda(\lambda,a) =
\sum_{n=0}^\infty \left.\frac{(\ap[_t] + \pp_t)^n(\lambda)}{(n+1)!}
\right|_{t=0}.
\label{SW2}
\en
Equations (\ref{SW2}) and (\ref{as}) with (\ref{flow}) are explicit
semi-classical versions of the Seiberg-Witten map.
The semi-classical (semi-noncommutative) generalized field strength
evaluated on two functions (e.g.\ coordinates) $f$, $g$  is
\eq
F_a(f,g) = \pcom{\rho^* f}{\rho^* g} - \rho^*\pcom{f}{g}
         = \rho^*\left(\pcom{f}{g}' - \pcom{f}{g}\right). \label{scgfs}
\en
Abstractly as 2-cochain:
\eq
F_a = \rho^* \circ \frac{1}{2}(\theta' -\theta)^{ik} \pp_i\wedge\pp_k
    = \rho^* \circ \frac{1}{2} (f')_{jl} \theta^{ij}\theta^{kl}\pp_i\wedge\pp_k
\en
with $\theta' f = \theta f'$, or
\eq
f' = \frac{1}{1 + f \theta} f,
\en
which we recognize as the  noncommutative field strength (with lower indices)
for constant $f$, $\theta$~\cite{SW}. The general result for non-constant $f$, $\theta$
is thus simply obtained by the application of the covariantizing map $\rho^*$
(after raising indices with $\theta$'s).

The Seiberg-Witten map in the semiclassical regime for constant $\theta$
has previously been discussed in~\cite{Cornalba1,Ishibashi}, where
it was understood as a coordinate
redefinition that eliminates fluctuations around a constant background.

We will now  use
Kontsevich's formality theorem to quantize everything. The goal is to
obtain (\ref{SWcond}) in the form (\ref{swcond})
of which (\ref{cswcond}) is the semi-classical limit.

\subsubsection{Kontsevich formality map}

Kontsevich's formality map 
is a collection  of skew-symmetric multilinear maps $U_n$ for $n=0 \ldots \infty$
that map tensor products of $n$ polyvector fields to differential operators.
More precisely
$U_n$ maps the tensor product
of $n$ $k_i$-vector fields to an $m$-differential operator, where $m$ is determined
by the matching condition
\eq
m = 2 - 2n + \sum_{i=1}^n k_i. \label{matching}
\en
$U_1$ in particular is the natural map from a $k$-vector field to a
$k$-differential operator
\eq
U_1(\xi_1\wedge\ldots\wedge\xi_k)(f_1,\ldots,f_k) = 
\frac{1}{k!}\sum_{\sigma\in\Sigma_k}
\mathrm{sgn}(\sigma) \prod_{i=1}^k \xi_{\sigma_i}(f_i),
\en
and $U_0$ is defined to be the ordinary multiplication of functions:
\eq
U_0(f,g) = fg.
\en
The $U_n$, $n \geq 1$, satisfy the formality condition~\cite{Kontsevich}
\eqa
\lefteqn{d_\mu U_n(\al_1,\ldots,\al_n) + \frac{1}{2}
\sum_{{}^{I\sqcup J = (1,\ldots,n)}_{I,J \neq \emptyset}} \pm
\gcom{U_{|I|}(\al_I)}{U_{|J|}(\al_J)}}&& \nn
&&=\sum_{i<j} \pm U_{n-1}\left(\sncom{\al_i}{\al_j},\al_1,\ldots,\widehat\al_i,
\ldots,\widehat\al_j,\ldots,\al_n\right),
\ena
where $d_\mu \CC \equiv -\gcom{\CC}{\mu}$, with the commutative multiplication $\mu(f,g)
= f\cdot g$
of functions; the hat  marks an omitted vector field.
See~\cite{Kontsevich,CattaneoFelder} for explicit constructions and more details
and~\cite{Arnal,Kontsevich} for the definition of the signs  in
this equation. In the following we collect the three
special cases that we actually use in this paper.

Consider the formal series (see also~\cite{Manchon})
\eq
\Phi(\alpha) = \sum_{n=0}^\infty \frac{(i\hbar)^n}{n!}
U_{n+1}(\alpha,\theta,\ldots,\theta).
\en
According to the matching condition (\ref{matching}),
$U_{n+1}(\alpha,\theta,\ldots,\theta)$ is a
tridifferential operator for every $n$ if $\alpha$ is a trivector field,
it is a bidifferential operator if $\alpha$ is a bivector field,
it is a differential operator if $\alpha$ is a vector field and
it is a function if $\alpha$ is a function; in all cases $\theta$
is assumed to be a bidifferential operator.

\paragraph{Star products from Poisson tensors}

A Poisson bivector $\theta$ gives rise to a star product via the formality map:
According to the matching condition (\ref{matching}),
$U_n(\theta,\ldots,\theta)$ is a
bidifferential operator for every $n$ if $\theta$ is a bivector field.
This can be used to define a product
\eq
f \star g  =  \sum_{n=0}^\infty \frac{\left(i\hbar\right)^n}{n!} 
U_n(\theta,\ldots,\theta)(f,g) 
 =  f g + \frac{i\hbar}{2}\,\theta^{ij}\pp_i f\,\pp_j g + \cdots \; .
\label{kstar}
\en
The formality condition
implies
\eq
\ds \star = i\hbar\Phi(\dpo \theta), \label{control}
\en
or, $\gcom{\star}{\star} = i\hbar\Phi(\sncom{\theta}{\theta})$,
i.e., 
associativity of $\star$,
if $\theta$ is Poisson.
(If $\theta$ is not Poisson, i.e., has non-vanishing
Schouten-Nijenhuis bracket $\sncom{\theta}{\theta}$,
then the product $\star$ is not associative, but the
non-associativity is nevertheless under control
via the formality condition by (\ref{control}).)

\paragraph{Differential operators from vector fields}

We can define a
linear differential operator\footnotemark\addtocounter{footnote}{-1}
\eq \label{vecdiff}
\Phi(\xi)
%\sum_{n=0}^\infty \frac{\left(i\hbar\right)^{n}}{n!}
%U_{n+1}(\xi,\theta,\ldots,\theta)
= \xi + \frac{(i\hbar)^2}{2}
U_3(\xi,\theta,\theta) + \cdots
\en
for every vector field $\xi$.
For $\theta$ Poisson the formality condition gives
\eq
\ds \Phi(\xi) = i\hbar \Phi(\dpo\xi)
%\sum_{n=0}^\infty \frac{\left(i\hbar\right)^{n+1}}{n!} 
%U_{n+1}(\dpo\xi,\theta,\ldots,\theta)
= i\hbar \,\dpo\xi + \cdots \; .
\label{dderiv}
\en
Vector fields $\xi$ that preserve the Poisson bracket, $\dpo\xi =
-\sncom{\theta}{\xi} = 0$, give rise to derivations of the star product
(\ref{kstar}): From (\ref{dderiv}) and the definition (\ref{hochschildd}),
(\ref{hochschilddc}) of $\ds$
\eq 
0 = [\ds\Phi(\xi)](f,g) = -[\Phi(\xi)](f\star g) + 
f \star [\Phi(\xi)](g) + [\Phi(\xi)](f)
\star g .
\en

\paragraph{Inner derivations from Hamiltonian vector fields}

Hamiltonian vector fields $\dpo f$ give rise to inner derivations of the
star product (\ref{kstar}):
%According to the matching condition 
%(\ref{matching}),
%$U_{n+1}(f,\theta,\ldots,\theta)$ is a
%function for every $n$ if $f$ is a function and $\theta$
%is a bidifferential operator.
We can define
a new function\footnote{$U_2(\xi,\theta) = 0$ and $U_2(f,\theta)=0$
by explicit computation of Kontsevich's formulas.}
\eq
\hat f \equiv \Phi(f) =
%\sum_{n=0}^\infty \frac{\left(i\hbar\right)^{n}}{n!}
%U_{n+1}(f,\theta,\ldots,\theta) =
f + \frac{(i\hbar)^2}{2}
U_3(f,\theta,\theta) + \cdots  \label{fhat}
\en
for every function $f$.
For $\theta$ Poisson the formality condition gives
\eq
\ds \hat f = i\hbar \Phi(\dpo f)
%\sum_{n=0}^\infty \frac{\left(i\hbar\right)^{n+1}}{n!}
%U_{n+1}(\dpo f,\theta,\ldots,\theta)
%= i\hbar\,\delta_{(\dpo f)}.
\label{dfhat}
\en
Evaluated on a function $g$, this reads
\eq
[\Phi(\dpo f)](g) = \frac{1}{i\hbar}\scom{g}{\hat f} . 
\en
The Hamiltonian vector field $\dpo f$ is thus mapped to the inner derivation
$\frac{i}{\hbar}\scom{\hat f}{\,\cdot\,}$.

\subsubsection{Quantum construction}

The construction mirrors the semiclassical one, the exact correspondence
is given by the formality maps $U_n$ that are skew-symmetric 
multilinear  maps that take  $n$ polyvector fields
into a polydifferential operator.  We start with the differential
operator 
\begin{equation}
  \label{astar}
  \as = \sum_{n=0}^\infty \frac{\left(i\hbar\right)^{n}}{n!}
U_{n+1}(\ap,\theta,\ldots,\theta),
\end{equation}
which is
the image of the vector field $\ap$ under the formality map 
(\ref{vecdiff}); then we use the coboundary operator 
$\ds$ (\ref{dq}) to define
a bidifferential operator
\begin{equation}
  \label{fstar}
  \fs = \ds\as.
\end{equation}
This is the image of $\fp = \dpo\ap$ under
the formality map:
\eq 
  \label{fstarformal}
  \fs = \sum_{n=0}^\infty 
  \frac{(i\hbar)^{n+1}}{n!}U_{n+1}(\fp,\theta,\ldots,\theta).
\en
A $t$-dependent Poisson structure (\ref{evolution}) induces a $t$-dependent
star product via (\ref{kstar})
\begin{equation}
  \label{tstar}
  g \star_t h = \sum_{n=0}^\infty
  \frac{(i\hbar)^{n}}{n!} U_n(\theta_t,\ldots,\theta_t)(g,h).
\end{equation}
The $t$-derivative of this equation is
\begin{equation}
  \label{tstarder}
  \pp_t(g \star_t h) =  \sum_{n=0}^\infty
  \frac{(i\hbar)^{n+1}}{n!} 
  U_{n+1}(\fp[_t],\theta_t,\ldots,\theta_t)(g,h),
\end{equation}
where we have used (\ref{evolution}) 
and the skew-symmetry and multi-linearity
of $U_n$.
Comparing with (\ref{fstarformal}) we find
\begin{equation}
  \label{tstarf}
  \pp_t(g\star_t h) = \fs[_t](g,h),
\end{equation}
or, shorter, as an operator equation: $\pp_t(\star_t) = \fs[_t]$. But
$\fs[_t] = \ds[_t] \as[_t] = -\gcom{\as[_t]}{\star_t}$,
so the $t$-evolution is generated by the differential operator
$\as[_t]$
and can be integrated to a flow (see appendix~\ref{s:t-evolution})
\begin{equation}
  \label{Dq}
  \DD_a = \left.\exp(\as[_t] + \pp_t)\exp(-\pp_t)\right|_{t=0} ,
\end{equation}
that relates the star products $\star' = \star_1$
and $\star = \star_0$, and that defines the generalized noncommutative
gauge potential
\begin{equation}
  \label{ncgaugepot}
  \AA_a = \DD_a - \id .
\end{equation}
The transformation of $\as$ under 
an infinitesimal gauge transformation $a \mapsto a +  d\lambda$
can be computed from (\ref{astrafo}) with the help of (\ref{dfhat}),
see (\ref{hochschilddonf}):
\begin{equation}
  \label{deltaaq}
  \as \mapsto \as + \frac{1}{i\hbar} \ds\hat\lambda .
\end{equation}
The effect of this transformation on the quantum flow and on the noncommutative
gauge potential are (see appendix~\ref{lambdatilde})
\eq
\DD_{a+d\lambda} = (\id +\frac{1}{i\hbar} \ds\hat\Lambda)\circ\DD_a,
\quad\mbox{i.e.,}\quad\DD_{a+d\lambda}(f) = 
\DD_a f + \frac{i}{\hbar}\scom{\hat\Lambda}{\DD_a f}
%- \DD_a f\star\tilde{\hat\lambda}\right)
\label{Dadl}
\en
\eq
\AA_{a+d\lambda} = \AA_a + \frac{1}{i\hbar}\left(\ds\hat\Lambda 
- \hat\Lambda\star\AA +\AA\star
\hat\Lambda\right).
\en
with
\eq
\hat\Lambda(\lambda,a) =
\sum_{n=0}^\infty \left.\frac{(\as[_t] + \partial_t)^n(\hat\lambda)}{(n+1)!}
\right|_{t=0}.
\label{tildehatlambda}
\en
Equations (\ref{ncgaugepot}) with (\ref{Dq}) and (\ref{tildehatlambda}) are
explicit versions of the abelian Seiberg-Witten map to all orders
in $\hbar$. They are unique up to
(noncommutative) gauge transformations.
Perhaps more importantly this construction provides us with
an explicit version of the ``covariantizer'' $\DD_a$ (the equivalence
map that sends coordinates and functions to their covariant
analogs) in terms of a finite number of (classical) fields
$a_i$.
The noncommutative gauge parameter (\ref{tildehatlambda}) also
satisfies the consistency condition
\eq
\delta_\alpha \hat\Lambda(\beta,a) - \delta_\beta \hat\Lambda(\alpha,a)
= \frac{i}{\hbar}\scom{\hat\Lambda(\alpha,a)}{\hat\Lambda(\beta,a)},
\en
with $\delta_\alpha (a_i) = \pp_i \alpha$, $\delta_\alpha(\beta) = 0$,
that follows from computing the commutator of abelian
gauge transformations on a covariant field~\cite{Jurco:2000ja}.

The generalized noncommutative field strength evaluated
on two functions (or coordinates) $f$, $g$ is
\eq
\FF_a(f,g) = \DD_a\left([f\stackrel{\star'}{,} g] - \scom{f}{g}\right).
\en

Up to order $\theta^2$ the series for $\AA_a$ and $\Lambda$
agree with the semiclassical results. In components:
\eq
\AA_a(x^i) = \theta^{ij} a_j + \frac{1}{2} \theta^{kl}a_l (\partial_k
(\theta^{ij} a_j) - \theta^{ij} f_{jk}) + \ldots ,
\en
\eq
\hat\Lambda = \lambda + \frac{1}{2} \theta^{ij} a_j \partial_i\lambda +
\frac{1}{6} \theta^{kl} a_l (\partial_k(\theta^{ij} a_j \partial_i \lambda)
-\theta^{ij} f_{jk} \partial_i \lambda) + \ldots .
\en

\noindent There are three major strategies for the computation of the Seiberg-Witten map:
\begin{enumerate}
\item[(i)] From the gauge equivalence condition (\ref{SWcond}) one can directly
obtain recursion relations for the terms in the Seiberg-Witten map.
For constant $\theta$ these can be cast in the form of differential
equations~\cite{SW}. Terms of low order in the gauge fields but all
orders in $\theta$ can be expressed in terms of
$\star_n$-products~\cite{Mehen:2000vs,LiuII}.\footnote{Some
motivation for the latter was provided
in~\cite{Das1}, and~\cite{Das2} provided some more concrete understanding of the
relationship of the generalized star product and Seiberg-Witten map.}

\item[(ii)] A path integral approach can be based on the relationship
between open Wilson lines in the
commutative and noncommutative picture~\cite{Okuyama,Okuyama:2001sw}.
\item[(iii)] The equivalence of the star products corresponding to
the perturbed and unperturbed Poisson structures leads to
our formulation in the framework of deformation quantization.
This allows a closed formula for the Seiberg-Witten map to
all orders in the gauge fields and in $\theta$.
\end{enumerate}

\subsection{Covariance and (non)uniqueness}
\label{sec:coordtransform}

The objects $\DD_a$ and $\hat\Lambda$ are not unique if all we ask is
that they satisfy the generalized Seiberg-Witten condition (\ref{Dadl})
with star product $\star$
and have the correct ``classical limit'' $\DD = \ap + \ldots$, $\hat\Lambda = \lambda
+ \ldots$.  The pair $\DD_2\circ\DD_a\circ\DD_1$, $\DD_2(\hat\Lambda)$,
where $\DD_2$ is an \mbox{$\star$-alge}\-bra automorphisms and $\DD_1$ is an equivalence map
(possibly combined with a change of coordinates of the
form $\id + o(\theta^2)$) is an equally valid solution. 
If we allow also a transformation to a new (but equivalent) star product 
$\overline\star$
then we can relax the condition on $\DD_2$: it may be any
fixed equivalence map possibly combined with a change of coordinates.
The maps (differential operators)
$\DD_1$ and $\DD_2$ may depend on the gauge potential $a_i$ via $f_{ij}$; it is 
important, however, that they are gauge-invariant.
The freedom in the choice of $\DD_1$ and $\DD_2$ represents the
freedom in the choice of coordinates and/or quantization scheme
in our construction; different $\DD_a$, $\hat\Lambda$
related by $\DD_1$, $\DD_2$ should be regarded as being equivalent.
\begin{figure}[tb]
\begin{center}
\begin{picture}(0,0)%
\includegraphics{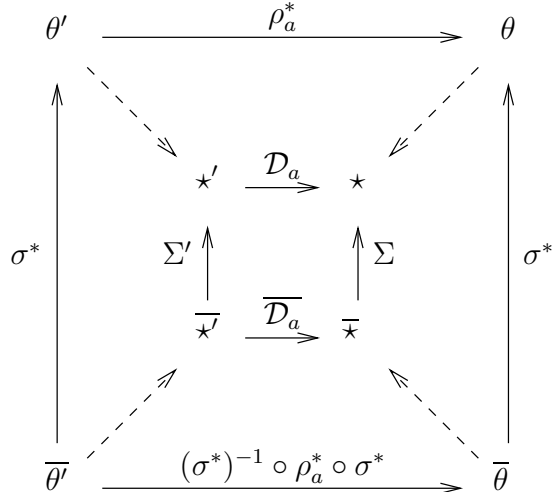}%
\end{picture}%
\setlength{\unitlength}{4144sp}%
\begingroup\makeatletter\ifx\SetFigFont\undefined%
\gdef\SetFigFont#1#2#3#4#5{%
  \reset@font\fontsize{#1}{#2pt}%
  \fontfamily{#3}\fontseries{#4}\fontshape{#5}%
  \selectfont}%
\fi\endgroup%
\begin{picture}(2880,2937)(811,-2773)
\put(3601,-61){\makebox(0,0)[c]{\smash{\SetFigFont{12}{14.4}{\rmdefault}{\mddefault}{\updefault}
$\theta$
}}}
\put(1801,-961){\makebox(0,0)[c]{\smash{\SetFigFont{12}{14.4}{\rmdefault}{\mddefault}{\updefault}
$\star'$
}}}
\put(2701,-961){\makebox(0,0)[c]{\smash{\SetFigFont{12}{14.4}{\rmdefault}{\mddefault}{\updefault}
$\star$
}}}
\put(1801,-1861){\makebox(0,0)[c]{\smash{\SetFigFont{12}{14.4}{\rmdefault}{\mddefault}{\updefault}
$\overline{\star'}$
}}}
\put(2656,-1861){\makebox(0,0)[c]{\smash{\SetFigFont{12}{14.4}{\rmdefault}{\mddefault}{\updefault}
$\overline{\star}$
}}}
\put(901,-2761){\makebox(0,0)[c]{\smash{\SetFigFont{12}{14.4}{\rmdefault}{\mddefault}{\updefault}
$\overline{\theta'}$
}}}
\put(3556,-2761){\makebox(0,0)[c]{\smash{\SetFigFont{12}{14.4}{\rmdefault}{\mddefault}{\updefault}
$\overline{\theta}$
}}}
\put(901,-61){\makebox(0,0)[c]{\smash{\SetFigFont{12}{14.4}{\rmdefault}{\mddefault}{\updefault}
$\theta'$
}}}
\put(2251, 29){\makebox(0,0)[b]{\smash{\SetFigFont{12}{14.4}{\rmdefault}{\mddefault}{\updefault}
$\rho^*_a$
}}}
\put(2251,-871){\makebox(0,0)[b]{\smash{\SetFigFont{12}{14.4}{\rmdefault}{\mddefault}{\updefault}
$\DD_a$
}}}
\put(2251,-1771){\makebox(0,0)[b]{\smash{\SetFigFont{12}{14.4}{\rmdefault}{\mddefault}{\updefault}
$\overline{\DD_a}$
}}}
\put(2251,-2671){\makebox(0,0)[b]{\smash{\SetFigFont{12}{14.4}{\rmdefault}{\mddefault}{\updefault}
$(\sigma^*)^{-1}\circ\rho_a^*\circ\sigma^*$
}}}
\put(1711,-1411){\makebox(0,0)[rt]{\smash{\SetFigFont{12}{14.4}{\rmdefault}{\mddefault}{\updefault}
$\Sigma'$
}}}
\put(2791,-1411){\makebox(0,0)[lt]{\smash{\SetFigFont{12}{14.4}{\rmdefault}{\mddefault}{\updefault}
$\Sigma$
}}}
\put(811,-1411){\makebox(0,0)[rt]{\smash{\SetFigFont{12}{14.4}{\rmdefault}{\mddefault}{\updefault}
$\sigma^*$
}}}
\put(3691,-1411){\makebox(0,0)[lt]{\smash{\SetFigFont{12}{14.4}{\rmdefault}{\mddefault}{\updefault}
$\sigma^*$
}}}
\end{picture}
\end{center}
\caption{Two nested commutative diagrams that illustrate
the covariance of the semiclassical and quantum constructions
under a change of coordinates given by $\sigma^*$.
The dashed lines indicate Kontsevich quantization.
$\Sigma$ and $\Sigma'$ are the equivalence maps (including $\sigma^*$)
that relate the star products that were computed in the new
coordinates with those computed in the old coordinates.
The top and bottom trapezia illustrate the construction of
the covariantizing equivalence maps in the old and new coordinates.} 
\label{pic:coordtransform}
\end{figure}

Figure~\ref{pic:coordtransform} illustrates how the semi-classical and
quantum  constructions are affected by a change of coordinates $\sigma^*$:
The quantization of $\overline{\theta}$ and $\overline{\theta'}$ in
the new coordinates leads to star products $\overline{\star}$ and
$\overline{\star'}$ that are related to the star products
$\star$ and $\star'$ in the old coordinates 
by equivalence maps $\Sigma$ and $\Sigma'$ respectively. (We have
included $\sigma^*$ in the definition of these maps.) Note that in general
$\Sigma \neq \Sigma'$.
The covariantizing
equivalence map, generalized gauge potential and field strength in the
new coordinates and old coordinates are related by:
\eqa
\overline{\DD_a} & = & \Sigma^{-1} \circ \DD_a \circ \Sigma' , \\
\overline{\AA_a} & = & \Sigma^{-1} \circ (\Sigma' - \Sigma) + \Sigma^{-1} \circ
\AA_a \circ \Sigma' , \\
\overline{\FF_a} & = & \Sigma^{-1} \circ \FF_a \circ (\Sigma')^{\otimes 2} .
\ena
Explicit (but complicated) expressions for
$\Sigma$ and $\Sigma'$ in terms of $\theta$, $\sigma^*$, and the
gauge potential $a$ can be computed with methods similar to the ones that
we have used to compute $\DD_a$ in the previous section.

\subsection{Dirac-Born-Infeld action in the intermediate picture}
\label{DBI}

Seiberg and Witten have argued that the open string theory effective
action in the presence of a background $B$-field can be expressed either in terms
of ordinary gauge theory written in terms of the combination $B+F$ or
in terms of noncommutative gauge theory with
gauge field $\widehat F$, where the $B$-dependence appears
only via the $\theta$-dependence of the star product and the open string
metric $G$ and effective coupling $G_s$. (Here we implicitly need to assume
that $\theta$ is Poisson, which is of course the case for constant $\theta$.)
There is also an intermediate picture 
with an effective noncommutative action which is a function of
$\widehat \Phi + \widehat F$, where $\widehat\Phi$ is a covariant version of
some antisymmetric matrix $\Phi$, with a $\theta$-dependent
star product and effective metric $G$ and string coupling $G_s$.
The proposed relations between
the new quantities and the background field $B$,
the given  closed string metric $g$ and the coupling $g_s$ are%
\footnote{To avoid confusion with the
matrices we will use bold face letters
for tensors and forms in this section
(e.g.: $\btheta = \frac{1}{2} \theta^{ij} \pp_i \wedge \pp_j$,
$\bomega = \frac{1}{2} \omega_{ij} dx^i \wedge dx^j$, $\theta = \omega^{-1}$);
for simplicity we shall assume that all matrices are nondegenerate when
needed.}
\newcommand*\deh{{\det}^{\frac{1}{2}}}
\newcommand*\dehi{{\det}^{-\frac{1}{2}}}
\eq
\frac{1}{G + \Phi} = \frac{1}{g + B} - \theta,
\qquad \frac{\deh(g + B)}{g_s} = \frac{\deh(G + \Phi)}{G_s}.
\en
The first relation can also be written more symmetrically:
\eq
[1+(G+\Phi)\theta][1-(g+B)\theta] = 1; \qquad G_s = g_s \dehi[1 - (g +
B)\theta]. \label{identities}
\en
Given $g$ and $B$ we can pick essentially \emph{any} antisymmetric matrix $\theta$
-- in particular one that satisfies the Jacobi identity
--
and find $G$ and $\Phi$ as symmetric and antisymmetric parts of the
following expression
\eq
G + \Phi = \frac{1}{1-(g+B)\theta} (g+B).
\en
For $\theta = 1/B$ (as in the zero slope limit):
$G = -B g^{-1} B$, $\Phi = -B$, $G_s = g_s \deh(-Bg^{-1})$.

For slowly varying but not necessarily small fields on a $D$-brane the
effective theory is given by the Dirac-Born-Infeld (DBI) action. In the following
we will show that the ordinary DBI action is exactly equal to the
semi-noncommutative DBI action in the intermediate picture. There are
no derivative corrections. By \emph{semi}-noncommutative we mean the
semiclassical limit of a noncommutative theory with star commutators,
e.g.\ in the noncommutative transformation law, replaced by Poisson brackets
as in section~\ref{sec:semiclass}.
Using (\ref{identities}) we can derive the following identity for scalar
densities
\eq
\frac{1}{g_s} \deh(g+ B + F) = \frac{1}{G_s}
\deh\!\left(\frac{\theta}{\theta'}\right) \deh(G+ \Phi + F'), \label{scalardens}
\en
where
\eq
\theta' = \theta\frac{1}{1 + F\theta},\qquad F' = \frac{1}{1+F\theta} F
\en
(understood as formal power series).
Raising indices on $F'$ with $\theta$ we get $\theta' - \theta
\equiv \pcom{x^i}{x^j}' - \pcom{x^i}{x^j}$ which we recognize as the
semiclassical version of
\eq
\tilde F^{ij} = [x^i\stackrel{\star'}{,} x^j] -
\scom{x^i}{x^j}, \label{ftild}
\en
compare equation (\ref{ftilde}). The semi-noncommutative field strength
changes by canonical transformation under gauge transformations (\ref{ctrafo}); 
it is obtained from the invariant $\theta' - \theta$ by action of the
covariantizing $\rho^*$  (see also (\ref{tensor}), (\ref{scgfs})): 
\eq
\rho^*(\theta' -\theta) = \pcom{\rho^* x^i}{\rho^* x^j} -
\rho^*\left(\pcom{x^i}{x^j}\right).
\en
The corresponding object with
lower indices is
\eq
\widehat F =\rho^* \left(F'\right).
\en 
The  Poisson structures $\btheta'$ and $\btheta$ are related by the change of
coordinates
$\rho^*$: $\rho^*\btheta' = \btheta$. The matrices $\theta'$, $\theta$
are consequently related to the Jacobian $\det(\pp \rho^*(x)/\pp x)$
of $\rho^*$:
\eq
\deh\!\left(\frac{\theta}{\rho^*\theta'}\right)\cdot\det\!\left(\frac{\pp
\rho^*(x)}{\pp x}\right) = 1.
\en
Using this we can derive the following exact equality for the DBI action with
background field $B$ and the semi-noncommutative DBI action without
$B$ (but with $\Phi$) in the intermediate picture,
\eq
\int d^px\,\frac{1}{g_s} \deh(g+ B + F) 
= \int d^px\,\frac{1}{\widehat G_s}
\frac{\deh(\rho^*\theta)}{\deh\theta}\deh(\widehat G + \widehat \Phi +
\widehat F), \label{sncdbi}
\en
with covariant $\widehat G_s \equiv \rho^* G_s$, $\widehat G \equiv \rho^* G$,
$\widehat \Phi \equiv \rho^* \Phi$, $\widehat F = \rho^* F'$ that transform
semi-classically under gauge transformations (\ref{ctrafo}).
The only object without a ``$\rho^*$'', $\dehi\theta$, is important
since it ensures that the semi-noncommutative action is invariant under
gauge transformations, i.e.\ canonical transformations.
The factor $\deh(\rho^*\theta)/\deh(\theta)$ can be absorbed in a redefinition
of $G_s$; it is equal to one in the case of constant $\theta$, since
$\rho^*$ does not change a constant;
\eq
\int d^px\,\frac{1}{g_s} \deh(g+ B + F) 
= \int d^px\,\frac{1}{\widehat G_s}
\deh(\widehat G + \widehat \Phi + \widehat F) \qquad\mbox{($\theta$ const.)} .
\en
General actions invariant under the semiclassical Seiberg-Witten map, which includes 
the Born-Infleld action as well as some actions with derivative terms,
have been discussed in~\cite{Cornalba2} in the case of constant
$\theta = B^{-1}$ and constant metric $g$.

One can consider a fully noncommutative version of
the DBI action with $\rho^*$ replaced by the equivalence map $\DD$ and with
star products
in the appropriate places.  That action no longer exactly equals
its commutative cousin but differs only by derivative terms as
can be seen~\cite{Paolo} by using the explicit form of the
equivalence map (\ref{Dq}) --
ordering ambiguities in the definition of the action
also contribute only derivative terms.
The equivalence up to derivative terms
of the commutative and noncommutative DBI actions was previously
shown  by direct computation~\cite{SW} in the case of constant $\theta$;
in this case an alternative derivation that is closer to 
our present discussion and is
based on a conjectured formula for the Seiberg-Witten map
was given in~\cite{LiuII}. Requireing equivalence of the
commutative and noncommutative descriptions one can compute
derivative corrections to the DBI action~\cite{Cornalba:2000ua}.

The semi-noncommutative actions have the general form
\eq
S_\Phi = \int d^px \, \frac{1}{\deh\theta}\,
\rho^*_{(\theta,a)}\Big(\mathcal{L}(G_s,G,\Phi,\theta',\theta)\Big),
\en
where $\mathcal{L}$ is a gauge invariant scalar function. The gauge potential
enters in two places: in $\theta'$
via the gauge invariant field strength $f$  and in
$\rho^*_{(\theta,a)}$. It is interesting to note that even the metric
and the coupling constants will in general transform under gauge
transformations since they depend on the gauge potential
via $\rho^*$.
Under gauge transformations $\rho^*\mathcal{L}$
transforms canonically:
\eq
\delta_\lambda (\rho^*\mathcal{L}) = \pcom{\rho^*\mathcal{L}}{\tilde\lambda}.
\en
Due to the special scalar density $\dehi\theta$, the action $S_\Phi$ is
gauge invariant and covariant under general coordinate transformations.

In the ``background independent'' gauge $\theta = B^{-1}$, $\Phi = -B$
\cite{Seiberg} the action (\ref{sncdbi}) becomes simply
\eq
 S_{\mathrm{DBI}} = \int d^px \, \frac{1}{\deh\theta}\: \rho^*\!\left(\frac{1}{g_s}
 \deh(1 + g \theta')\right),
\en
with $\theta' = (B+F)^{-1}$.
Expanding the determinant to lowest nontrivial order we find the following
semi-non\-com\-mu\-ta\-tive Yang-Mills action
\eq
\int d^px \, \frac{1}{\deh\theta}\: \rho^*\!\!\left(\frac{1}{4 g_s}  g_{ij}
\theta'^{jk} g_{kl} \theta'^{li}\right) =
\int d^px \, \frac{1}{\deh\theta}\, \frac{1}{4\hat g_s} \hat g_{ij}
\pcom{\hat x^j}{\hat x^k} \hat g_{kl} \pcom{\hat x^l}{\hat x^i},
\en
with covariant coupling constant $\hat g_s = \rho^* g_s$,
metric $\hat g_{ij} = \rho^* g_{ij}$ and coordinates $\hat x^i = \rho^* x^i$.
An analogous fully noncommutative version can be written with the
help of the covariantizing equivalence map $\DD$
\eq
S_{\mathrm{NC}} = \int d^px \, \deh\omega\: \DD\!\left(\frac{1}{4 g_s} \star' g_{ij} \star'
\tilde F^{jk} \star' g_{kl} \star' \tilde F^{li}\right),
\en
with an appropriate scalar
density $\deh\omega$ that ensures upon integration
a cyclic trace (this is important for the gauge invariance of the action.)
In the zero-slope limit $\tilde F$ is as given
in (\ref{ftild}), in the background independent gauge
$\tilde F^{ij} =
[x^i  \stackrel{\star'}{,}  x^j  ]$.
In the latter case
\eq
S_{\mathrm{NC}} = \int d^px \, \deh\omega\: \frac{1}{4 \hat g_s} \star \hat g_{ij} \star
\scom{\hat X^j}{\hat X^k} \star \hat g_{kl} \star \scom{\hat X^l}{\hat X^i}.
\en
This has the form of a matrix model potential (albeit with nonconstant
$g_s$, $g$) with covariant coordinates $\hat X^i = \DD x^i$ as dynamical variables.

\subsection{Some notes on symmetric tensors}
\label{metric}
The matrix
\eq
\wtheta^{ij} = \left(\frac{1}{B + g}\right)^{ij} = \theta^{ij} +
G^{ij} \label{matrix}
\en
plays a central role for strings in a background $B$-field (with $\Phi = 0$):
its symmetric part $\wtheta_S^{ij}$  is the effective open string
metric and its antisymmetric part $\wtheta_A^{ij}$
provides the Poisson structure (provided it is indeed Poisson)
that leads upon quantization to the noncommutativity
felt by the open strings. One may now ask out of pure curiosity whether
it is possible to quantize $\wtheta$ directly, i.e., whether it is
possible to find an associative
star product $\tilde\star$ which in lowest order in $\hbar$
is given by $\wtheta$ (which is not antisymmetric):
\eq
f \tilde\star g =  f g + \frac{i\hbar}{2} \wtheta^{ij} \pp_i f \pp_j g +
o(\hbar^2).
\en
This is indeed possible, provided $\wtheta_A$ is Poisson (i.e. satisfies
the Jacobi identity), since the symmetric part of the star product can
be gauged away by an equivalence map
\eq
\Xi(f\tilde\star g) = \Xi(f) \star \Xi(g)
\en
where
\eq
f \star g =  f g + \frac{i\hbar}{2} \wtheta_A^{ij} \pp_i f \pp_j g +
o(\hbar^2).
\en
An equivalence map that does the job can be given explicitly in terms
of the symmetric part of $\wtheta$:
\eq
\Xi = \exp(-\frac{i\hbar}{4} \wtheta_S^{ij} \pp_i \pp_j).
\en
It is enough to check  terms up to order $\hbar$
\eqa
\lefteqn{f g + \frac{i\hbar}{2}\wtheta^{ij}\pp_i f\pp_j g
-\frac{i\hbar}{4}\wtheta_S^{ij} \pp_i \pp_j (fg)}\nn
& = &f g + \frac{i\hbar}{2} \wtheta_A^{ij}\pp_i f\pp_j g
- \frac{i\hbar}{4} \wtheta_S^{ij}\left[(\pp_i\pp_j f) g + f (\pp_i\pp_j
g)\right]. \nonumber
\ena
To quantize $\wtheta$ we can thus proceed as follows: first one quantizes
$\wtheta_A$ e.g.\ with Kontsevich's formula and then one uses $\Xi$ to
get $\tilde\star$ from $\star$. In the previous sections we saw that
the full information about the noncommutative gauge fields is encoded
in the equivalence map $\DD_a$. Here we can similarly reconstruct the metric
field from $\Xi$ by evaluating its Hochschild field strength
\eq
\FF^\Xi_H(x^i,x^j) = \Xi(x^i) \star \Xi(x^j) - \Xi(x^i \star x^j)
= \frac{i\hbar}{2} G^{ij} + o(\hbar^2),
\en
or more directly: $G^{ij} = \frac{2i}{\hbar}\left(\Xi(x^i x^j) - x^i x^j\right)$.
Two more questions come up naturally: When is $\theta = (B + g)^{-1}_A$ Poisson?
Why is the relevant star product not a quantization of
$\tau \equiv B^{-1}$ as in the zero-slope limit? The answer to the second
question is of course that this is determined by the open string
propagator in the presence of a background $B$ and that happens to have
an antisymmetric part given by $\theta^{ij}$ and only in the zero-slope
limit this is equal to $B^{-1}$. Nevertheless the two
questions turn out to be related -- the star products based
on $\tau$ and $\theta$ are equivalent provided
that the 2-forms
\eq
\boldsymbol{B} = \frac{1}{2}(\tau^{-1})_{ij} dx^i\wedge dx^j, \qquad
\boldsymbol{\phi} = -\frac{1}{2}(g \tau g)_{ij} dx^i\wedge dx^j
\en
are closed, $\boldsymbol{\phi} = d\boldsymbol{\alpha}$ for
some 1-form $\boldsymbol{\alpha}$,  and $B(t) = B + t\phi$
is nondegenerate for $t\in[0,1]$; the closedness conditions
on $\boldsymbol{B}$ and $\boldsymbol{\phi}$
ensure in particular that $\alpha$ and $\theta$ are Poisson.
According to Moser's lemma the symplectic structures $B(1)$ and $B(0)$
are related by a change of coordinates generated by the vector
field $\chi_\alpha = \tau^{ij} \alpha_j \pp_i$
and, moreover, according to Kontsevich the star products resulting from
quantization of $B(0)$ and $B(1)$ are equivalent. Since $B(0) = \tau^{-1}$
and $B(1) = \theta^{-1}$ we have demonstrated our claim.
Let us remark that $\boldsymbol{B}$ and $\boldsymbol{\phi}$ are
closed if $g$ and $B$ are derived from a K\"ahler metric,
since then $\boldsymbol{B}$ is closed and $B$, $g B^{-1} g$ are proportional.
Instead of using Moser's lemma we could also drop some assumptions
and work directly with Poisson structures as in previous sections.
We would then require that $\tau$ is Poisson
and $g \tau g$ is closed and 
introduce a 1-parameter deformation $\tau(t)$, $t \in [0,1]$, with
\eq
\tau(0) = \tau, \qquad \pp_t \tau(t) = \tau(t)\cdot (g \tau g)\cdot \tau(t)
\en
and solution
\eq
\tau(t) = \tau + t \tau g \tau g \tau + t^2 \tau g \tau g \tau g \tau g \tau + \ldots = \tau\frac{1}{1 - t(g\tau)^2}.
\en
This is the antisymmetric part of
\eq
\wtheta(t) = \frac{1}{B + t^{\frac{1}{2}} g}.
\en
The symmetric part is $t^{\frac{1}{2}} G^{ij}(t) = -t^{\frac{1}{2}}
 [\tau(t) g \tau]^{ij}$
with $G(0) = -B^{-1} g B^{-1}$ and $G(1) = G$,
while $\tau(0) = B^{-1}$ and $\tau(1) = \theta$ with
$G$ and $\theta$ as given in (\ref{matrix}).
This suggest that $\phi = -g \tau g$ represents ``metric fluctuations''
around the background $B$ that can be gauged away by an equivalence
transformation that curiously leads to the zero-slope values $G(0)$, $\tau(0)$
of the metric and Poisson structure.

\section{Seiberg-Witten map for nonabelian gauge fields}

We will now extend the discussion to nonabelian gauge theories, i.e.,
Lie algebra-valued gauge potentials and gauge fields. We will argue
that a Seiberg-Witten map can be explicitly constructed for \emph{any}
gauge group by treating both the space-time noncommutativity and
the noncommutativity of the nonabelian gauge group on equal
footing. Both structures are obtained from appropriate Poisson structures
by deformation quantization. This construction
generalizes to fairly arbitrary noncommutative
internal spaces.

Let us mention that  it is possible to  absorb
a matrix factor (e.g. GL(n) or U(n) in the defining representation)
 directly into the definition of the noncommutative space $\Ax$
and then work with the abelian results of the previous sections,
however, for other gauge groups it is not a priory clear how to do this consistently.
In any case even for GL(n) and U(n) that approach would not give a very
detailed description of the nonabelian Seiberg-Witten map.

\subsection{Nonabelian setting}

In this section we shall establish notation and will
give a precise definition of the problem that
we would like to solve. Consider a manifold ``(noncommutative) space-time'' with
a noncommutative structure provided by a star product that is
derived from a Poisson structure $\Theta^{\mu\nu}$. On this space consider
a nonabelian gauge theory with gauge group G,
field strength $F_{\mu\nu}$, that can be locally expressed as
\eq
F_{\mu\nu} = \pp_\mu A_\nu - \pp_\nu A_\mu - i [A_\mu,A_\nu] 
\en
with nonabelian gauge potential $A_\mu = A_{\mu b} T^b$
where $T^b \in \mathrm{Lie}(G)$ are generators with
commutation relations
$-i[T^a,T^b] = C^{ab}_c T^c$, and nonabelian gauge transformations
\eq
\delta_\Lambda A_\mu = \pp_\mu \Lambda + i [\Lambda, A_\mu] .
\en
Our main goal is to find a noncommutative gauge potential $\As = \As(A_\mu)$ and
a noncommutative gauge parameter $\LLq(A_\mu,\Lambda)$  such that
a nonabelian gauge transformation $\delta_\Lambda$ of $A_\mu$ induces
a noncommutative gauge transformation $\deltaq_{\LLq}$ of $\As$:
\eq
\As(A_\mu + \delta_\Lambda A_\mu) = \As(A_\mu) + \deltaq_{\LLq} \As(A_\mu) .
\en
$\As(A_\mu)$ should be a universal enveloping algebra-valued
formal power series in $\Theta^{\mu\nu}$,
starting with $\Theta^{\mu\nu} A_\nu \pp_\mu$, that
contains polynomials of  $A_\mu$ and it's derivatives.
Similarly $\LLq(A_\mu,\Lambda)$ should be a universal enveloping algebra-valued
formal power series in $\Theta^{\mu\nu}$,
starting with $\Lambda$, that
contains polynomials of  $A_\mu$, $\Lambda$ and their derivatives.
The product in the definition of the noncommutative gauge transformation is
a combination of the star product on space-time and the matrix product of
the $T^a$.
We expect that it should be possible to find expressions, where the
structure constants $C^{ab}_c$ do not appear explicitly, except via commutators of
the Lie algebra-valued $A_\mu$, $\Lambda$.

A secondary goal is to find a construction that stays as close as possible to
the method that we used in the abelian case. There, we used a generalization of
Moser's lemma to relate Poisson structures $\btheta$ and $\btheta'$ (and, after
quantization, star products $\star$ and $\star'$). The motivation for this
and some of the complications of the nonabelian case can be most easily
understood in the special case of invertible, i.e. symplectic, Poisson structures.
The inverses of $\theta$ and $\theta'$ define closed 2-forms $\boldsymbol B$ and
$\boldsymbol B'$ that
differ by the addition of a (closed) gauge field $\boldsymbol f$ (\ref{BplusF}). Physically
$\boldsymbol B'$ is the background $\boldsymbol B$-field plus fluctuations
$\boldsymbol f$. In the nonabelian
case we would like to keep this picture but with $\boldsymbol f$ replaced by the
nonabelian field strength $\boldsymbol F$:
\[
{\boldsymbol B'} = {\boldsymbol B} + {\boldsymbol F} ,
\]
where $B = \Theta^{-1}$.
The trouble with this is that $d{\boldsymbol F} = -{\boldsymbol A}\wedge {\boldsymbol A}
\neq 0$ in the nonabelian case
so $\boldsymbol B$ and $\boldsymbol B'$ cannot both be closed 2-forms,
which they should be
if we want to interpret their inverses as Poisson structures. Ignoring this,
we could then look for a ``vector field'' $\chi$
that generates a coordinate transformation
that relates $\boldsymbol B$ and $\boldsymbol B'$. A natural generalization from the abelian
case (\ref{vf}) is
\[
\chi = \Theta^{\mu\nu} A_\nu D_\mu,
\]
where we have replaced the abelian gauge potential by the nonabelian one
and have also switched to a covariant derivative
$D_\mu = \pp_\mu + i[A_\mu , \,\cdot\,]$. The trouble here is that
it is not clear how to act with the matrix-valued $\chi$ on ``coordinates''.
$\chi$ is certainly no vector field; it is not even a derivation. (It's action
turns out to involve complete symmetrization over the constituent matrices
$T^a$.)

The solution to both problems is to consider a larger space that is spanned
by the space-time coordinates $x^\mu$ and by symbols $t^a$ for the
generators $T^a$ of the Lie Group. $\chi$~is then the projection onto
space-time of a true vector field and $\Theta$ and $\Theta'$ are 
the space-time components of true Poisson structures on the enlarged space.
We obtain the desired nonabelian noncommutative gauge theory by quantizing
both the external and internal part of the enlarged space at the same time.
To use the method of the previous sections we need to encode the
nonabelian data in an abelian gauge theory on the enlarged space. This program
is successful, if the  ``commutative'' nonabelian gauge theory can be recovered at
an intermediate step.

\subsection{Abelian data}

\emph{Notation:} Greek indices $\mu$, $\nu$, $\xi$, \ldots\ belong to
the external space, indices from the beginning of the alphabet $a$, $b$, $c$,
\ldots\ belong to the internal space and indices $i$, $j$, $k$,~\ldots\
run over the whole space (internal and external). We shall use
capital letters ($A_\mu$, $F_{\mu\nu}$, $\Theta^{\mu\nu}$) for things
related to the nonabelian theory on the external space and
small letters ($a_i$, $f_{ij}$, $\theta^{ij}$) for objects related to
the abelian theory on the enlarged space or the internal space
($a_b$, $\vartheta^{bc}$).

The $t^a$ are commutating coordinate functions on the internal space (``Lie algebra'') 
just like the $x^\mu$ are commuting coordinate functions on the external space
(``space-time''). We later recover the Lie algebra in the form of star-commutators
on the internal space and the matrices $T^a$ by taking a representation of that
algebra. The star product on the internal space is a quantization of
its natural Poisson structure
\eq
\Liecom{t^a}{t^b}  = C^{ab}_c t^c =: \vartheta^{ab}.
\en
In the new language
\eq
F_{\mu\nu} = \pp_\mu A_\nu - \pp_\nu A_\mu + \Liecom{A_\mu}{A_\nu} 
\en
with $A_\nu(x,t) = A_{\nu b}(x) t^b$ and
\eq
\delta_\Lambda A_\mu = \pp_\mu \Lambda + \Liecom{A_\mu}{\Lambda}
\en
with $\Lambda(x,t) = \Lambda_b(x) t^b$. (``Lie algebra-valued'' translates
into ``linear in $t$''.)
We equip the enlarged space with
a Poisson structure $\theta^{ij}$ which is the direct sum of the
external $\Theta^{\mu\nu}$ and internal $\vartheta^{ab}$ Poisson structures
\eq
\theta=
\left(\begin{array}{c|c} \Theta &  0 \\
\hline  0 & \vartheta \end{array}\right).
\en
Only for $t=0$ is the Poisson structure block-diagonal.
$\theta(t)$ and in particular $\theta' = \theta(1)$ acquire off-diagonal
terms through the $t$-evolution
\eq
\theta(0) = \theta,\; \pp_t \theta(t) = -\theta(t) f \theta(t),\mbox{ i.e., }
\theta(t) = \theta - t \theta f(t) \theta + t^2 \theta f(t) \theta f(t) \theta
\mp \cdots
\en
generated by an abelian gauge
field $f$ that is itself not block-diagonal (but whose internal
components $f_{ab}$ are zero as we shall argue below.)
The space-time components of $\theta(t)$ can be re-summed in a series
in $\Theta$ and we miraculously obtain an expression
that looks like the series for $\theta(t)$ but with
$f$ replaced by $F_{\mu\nu}(t) \equiv f_{\mu\nu} - t f_{\mu a} \theta^{ab}
f_{b\nu}$, which at $t=1$ (and $a_b = 0$, see below) becomes
the nonabelian field strength $F \equiv F(1) = 
\pp_\mu a_\nu - \pp_\nu a_\mu + \Liecom{a_\mu}{a_\nu}$:
\eqa
\Theta^{\mu\nu}(t) \, \equiv \, \theta^{\mu\nu}(t) &  = & \theta^{\mu\nu} - t\theta^{\mu i}f_{ij}\theta^{j\nu}
+ t^2 \theta^{\mu j} f_{jk} \theta^{kl} f_{lm} \theta^{m\nu} \mp \cdots \nn
& = & \theta^{\mu\nu} - t \theta^{\mu\kappa} \left(f_{\kappa\sigma} -
t f_{\kappa a} \theta^{ab} f_{b\sigma}\right) \theta^{\sigma\nu} + \cdots .
\ena
To all orders in $\Theta$:
\eq
\Theta(t)\, =\, \Theta - t \Theta F(t) \Theta + t^2 \Theta F(t) \Theta F(t) \Theta
\mp \cdots \,= \,\Theta\frac{1}{1 + t F(t) \Theta}.
\en
In the case of invertible $\Theta$, $\Theta'$ (with $\Theta' \equiv \Theta(1)$) we have
\eq
\frac{1}{\Theta'} = \frac{1}{\Theta} + F.
\en
This resembles the relation $B' = B + F$ that one would have naively
expected, but we should note that $\Theta$, $\Theta'$ are not necessarily
Poisson (they are just the space-time components of the Poisson
structures $\theta$, $\theta'$) and $F$ is not exactly a non-abelian
field strength (it is in fact a gauge-invariant expression in
abelian gauge fields that coincides with the nonabelian
field strength in the special gauge $a_b = 0$; see next section.) The other components of $\theta(t)$ are
computed similarly (again using $f_{ab} = 0$):
\eq
\theta^{\mu b}(t) = -\theta^{b
\mu}(t) = -t \Theta^{\mu\nu}(t) f_{\nu a} \vartheta^{ab},\qquad
\theta^{ab}(t) = \vartheta^{ab} + t^2 [\vartheta f \Theta(t) f
\vartheta]^{ab}.
\en
$\Theta(t)$ is not the only object that acquires a non-abelian look at $t = 1$;
this is also the case for Moser's vector field
\eq
\ap['] = (\theta')^{ij} a_j \pp_i = (\Theta')^{\mu\nu} \bar A_\nu \bar D_\mu +
\vartheta^{ab} a_b \pp_a,
\en
where $\bar A_\nu = a_\nu - f_{\nu a} \vartheta^{ab} a_b$ and
$\bar D_\mu = \pp_\mu - f_{\mu a} \vartheta^{ab} \pp_b$ are
the first terms in the expansions for the nonabelian gauge potential
and covariant derivative, valid around the special gauge $a_b = 0$ (see next
section.)

Now we need to identify appropriate abelian gauge fields and gauge
transformations on the
enlarged space that upon quantization give the desired
nonabelian noncommutative gauge fields and noncommutative gauge transformations.
For this we consider the terms of lowest order in $\Theta$ of
the Seiberg-Witten condition, where we expect
to see a purely nonabelian gauge transformation. Up to this order it is in fact enough to
work with the semiclassical condition (\ref{cswcond})
\[
A_{a+d\lambda} = A_a + \dpo\tilde\lambda + \{A_a,\tilde\lambda\}.
\]
Evaluating this on $x^\mu$ and collecting terms of order $\Theta$ we
get
\eq
A_\mu(a + d\lambda) = A_\mu(a) + \pp_\mu\Lambda + \Liecom{A_\mu(a)}{\Lambda},
\en
with $A_\mu(a)$ and $\Lambda(\lambda,a)$ defined
by
\eq
A_a(x^\nu) = \Theta^{\nu\mu} A_\mu(a) + o(\Theta^2),
\qquad \tilde\lambda = \Lambda(\lambda,a) + o(\Theta). \label{nadata}
\en
An abelian gauge transformation $\delta a_i = \pp_i \lambda$ thus results
in a nonabelian gauge transformation of $A_\mu$ with gauge parameter
$\Lambda$. Since we would like to identify $A_\mu$ and $\Lambda$
with the gauge potential and parameter of the ordinary nonabelian gauge theory
that we started with, they should both be linear in the coordinates $t^a$
of the internal space. Studying gauge transformations and
the explicit expression (\ref{flow}), (\ref{as})
for $A_a(x^\mu)$ (see next section) we find that this implies that the internal components
of the abelian gauge potential are independent of the $t^a$,
while the external components are linear in the $t^a$. This is preserved by
abelian gauge transformations with gauge parameters that are linear
in the $t^a$:
\eq
a_\mu = a_{\mu b}(x) t^b, \quad a_b = a_b(x),\quad \lambda
= \lambda_{b} (x) t^b; \quad \delta a_\mu = i(\pp_\mu \lambda_{b}) t^b,
\quad \delta a_b = i \lambda_{b}. \label{abdata}
\en
The gauge invariant characterization of the desired abelian
gauge fields is $f_{ab} = 0$, $f_{\mu b} = - f_{b\mu}$
independent of the~$t^a$, $f_{\mu\nu}$ linear in the $t^a$.
By a gauge transformation with parameter $-a_b(x) t^b$ we
can always go to a special gauge with vanishing internal gauge potential
$a_b' = 0$ (and $a'_\mu = a_\mu - \pp_\mu(a_b) t^b$.)

We can now apply the method that we developed for the abelian case
in the previous sections to obtain the desired nonabelian
noncommutative gauge fields in terms of the abelian data (\ref{abdata}).
These are $\theta$-expanded noncommutative gauge fields
that become ordinary nonabelian gauge fields at lowest nontrivial
order in $\Theta$ (but all orders in $\vartheta$.)
We claim that a re-summation of the $\theta$-series gives
in fact $\Theta$-expanded noncommutative gauge fields in terms
of nonabelian gauge fields. This is a much stronger statement
and can be checked by inspection using the special gauge $a_b = 0$.
A rigorous formal proof is however missing.

\subsection{Mini Seiberg-Witten map}

Nonabelian gauge theory is of course also a type of noncommutative
gauge theory and one may thus wonder whether a Seiberg-Witten map
exists from abelian to nonabelian gauge fields. Computing
$A_\mu(a)$ and $\Lambda(\lambda,a)$ (\ref{nadata}) using
the results from the previous sections does in fact provide
such maps:
\eq
A_\mu(a)  =  \left(e^{\api}\right)(a_\mu - \pp_\mu \alpha) +
\left(\frac{e^{\api} -1}{\api}\right)(\pp_\mu\alpha),
\en
\eq
\Lambda(\lambda, a)  =  \left(\frac{e^{\api} -1}{\api}\right)(\lambda)
\en
with $\alpha(x,t) = a_b(x) t^b$, the parameter of the 
gauge transformation that gives $a_b = \pp_b \alpha$ starting
from the special gauge $a_b = 0$. Note that $\api = \vartheta^{ab} a_b \pp_a =
\Liecom{\cdot}{\alpha}$. In components
\eq
A_\mu(a) = \sum_{n=0}^\infty \frac{1}{(n+1)!} t^a (M^n)^b_a(a_{\mu b} - n f_{\mu b}),
\en
\eq
\Lambda(\lambda,a) = \sum_{n=0}^\infty \frac{1}{(n+1)!} t^a (M^n)^b_a \lambda_b,
\en
with the matrix $M_a^b = C^{bc}_a a_c$ and
$a_{\mu b}t^b = a_\mu$, $\lambda_b t^b = \lambda$.
Under an abelian gauge transformation $\delta_\lambda a_i = \pp_i \lambda$,
\eq
\delta_\lambda A_\mu(a) = \pp_\mu \Lambda(\lambda, a)
+ \Liecom{A_\mu(a)}{\Lambda(\lambda, a)}.
\en
In the special gauge of vanishing internal gauge potential $a_b = 0$ the
maps
becomes simply $A_\mu(a) = a_\mu$, $\Lambda(\lambda,a) = \lambda$.

\section*{Acknowledgments}

We would like to thank P.~Aschieri, J.~Madore, S.~Schraml and A.~Sitarz for
helpful discussions.

\appendix

\renewcommand{\theequation}{\thesection.\arabic{equation}}
\newcommand{\newsection}{ \setcounter{equation}{0} \section}

\newsection{Brackets, evolution and parameters}

\subsection[Schouten-Nijenhuis bracket]{Schouten-Nijenhuis bracket\protect\footnote{%
A good reference for the material in this section and the next
is~\cite{Weinstein}.}}
\label{App1}

The Schouten-Nijenhuis bracket 
of two polyvector fields is defined
by
\eqa
\lefteqn{\sncom{\xi_1\wedge\ldots\wedge\xi_k}{\eta_1\wedge\ldots\wedge\eta_l}}
\phantom{\sncom{f}{\xi_1\wedge\ldots\wedge\xi_k}}&& \nn[2pt]
& = & \sum_{i,j}(-)^{i+j} \,
[\xi_i,\eta_j]\wedge\xi_1\wedge\ldots\wedge\hat{\xi}_i      %\nn &&\qquad
\wedge\ldots\wedge
\hat{\eta}_j\wedge\ldots\wedge\eta_l , \nn[5pt]
\sncom{f}{\xi_1\wedge\ldots\wedge\xi_k} &=&
\sum_{i}(-)^{i-1}\, \xi_i(f)\xi_1\wedge\ldots\wedge\hat{\xi}_i
\wedge\ldots\wedge\xi_k, \label{sncom}
\ena
if all $\xi$'s and $\eta$'s are vector fields and $f$ is a function.
The hat marks omitted vector fields.
A Poison tensor is a bivector field $\theta = 
\frac{1}{2}\theta^{ij}\pp_i\wedge\pp_j$
that satisfies the Jacobi identity
\eq
0 = \sncom{\theta}{\theta} \equiv 
\frac{1}{3}\Big(\theta^{il}\pp_l(\theta^{jk})
+\theta^{jl}\pp_l(\theta^{ki})
+\theta^{kl}\pp_l(\theta^{ij}) \Big)\pp_i\wedge\pp_j\wedge\pp_k .
\label{jacobi}
\en
In terms of the coboundary operator
\eq
\dpo = -\sncom{\cdot}{\theta},
\en
this can also be expressed as $\dpo\theta = 0$ or $\dpo^2 = 0$.

\subsection{Gerstenhaber bracket}
\label{App2}
 The Gerstenhaber bracket is
given by
\eq
\gcom{\CC_1}{\CC_2} = \CC_1 \circ \CC_2 - (-)^{(p_2+1)(p_1+1)} \CC_2 \circ \CC_1,
\label{gerstenhaber}
\en
where composition $\circ$
for $\CC_1 \in C^{p_1}$ and $\CC_2 \in C^{p_2}$ is defined as
\eqa
\lefteqn{(\CC_1 \circ \CC_2)(f_1,f_2,\ldots,f_{p_1+p_2-1})
= \CC_1\Big(\CC_2(f_1,\ldots,f_{p_2}),f_{p_2+1},\ldots,f_{p_2+p_1-1}\Big)}&&\nn
&& {}-(-)^{p_2} \CC_1\Big(f_1,
   \CC_2(f_2,\ldots,f_{p_2+1}),f_{p_2+2},\ldots,f_{p_2+p_1-1}\Big) \nn
&&   {}\pm \ldots + (-)^{(p_2+1)(p_1+1)}
   \CC_1\Big(f_1,\ldots,f_{p_1-1},\CC_2(f_{p_1},\ldots,f_{p_1+p_2-1})\Big) ;
\ena
$C^p$ may be either $\Hom_k(\Ax^{\otimes p},\Ax)$
or the space of $p$-differential operators $D_\mathrm{poly}^p$. 
In analogy to (\ref{jacobi}) we can express the 
associativity of a product $\star \in C^2$ as
\eq
\gcom{\star}{\star} = 0 , \qquad 
\gcom{\star}{\star}(f,g,h) \equiv 2\Big((f\star g)\star h - f\star(g\star
h)\Big).
\en
In terms of the coboundary operator (see also (\ref{hochschilddc}))
\eq
\ds: C^p \rightarrow C^{p+1}, \qquad \ds\CC = - \gcom{\CC}{\star},
\label{hochschildd}
\en
this can also be written as $\ds\star = 0$ or $\ds^2 = 0$.

\subsection{t-evolution}
\label{s:t-evolution}

Consider a $t$-dependent function $f(t)$ whose $t$-evolution is governed
by
\eq
\left(\pp_t + A(t)\right) f(t) = 0, \label{tevolution}
\en
where $A(t)$ is an operator (vector field or differential operator of arbitrary
degree) whose $t$-dependence is given. We are interested to relate $f(1)$ to
$f(0)$. There is a simple way to integrate (\ref{tevolution}) without having
to resort to $t$-ordered exponentials: By Taylor expansion
\eq
e^{-\pp_t} f(t) = f(t-1).
\en
Due to (\ref{tevolution}) we can insert $\exp(\pp_t + A(t))$ without
changing anything
\eq
e^{-\pp_t} e^{\pp_t + A(t)} f(t) = f(t-1).
\en
The trick hereby is that due to the Baker-Campbell-Hausdorff formula
all $\pp_t$ are saturated in the
product of the exponentials; there are no free $t$-derivatives
acting on $f(t)$, so we can evaluate at $t=1$ and get
\eq
\left.e^{-\pp_t} e^{\pp_t + A(t)}\right|_{t=1} f(1) = f(0) ,
\en
or, slightly rearranged
\eq
\left.e^{\pp_t + A(t)} e^{-\pp_t} \right|_{t=0} f(1) = f(0) .
\en
The first few terms in the expansion of the exponentials are
$1 + A + \frac{1}{2}(A^2 + \dot A) + \cdots$ .

\subsection{Semi-classical and quantum gauge parameters}
\label{lambdatilde}

Let $A$ and $B$ be two operators (vector fields or differential operators) and
\eq
B_0 \equiv B, \qquad B_{n+1} = [A , B_n],
\en
then
\eq
e^{A + \epsilon B} - e^A
= \epsilon\sum_{n=0}^\infty \frac{B_n }{(n+1)!} \: e^A
+ o(\epsilon^2) . \label{useful}
%\equiv \left(\frac{e^{[A,\cdot] - \id}}{[A,\cdot]}\right)(B) e^A + o(\epsilon^2)
\en
\paragraph{Semi-classical:}
We would like to proof (\ref{flow}) and (\ref{SW2})
\eq
\rho^*_{a+d\lambda} - \rho^*_a = (\dpo\tilde\lambda)\circ\rho^*_a + o(\lambda^2),
\qquad \tilde\lambda(\lambda,a) =
\sum_{n=0}^\infty \frac{\left.(\ap[_t] + \pp_t)^n(\lambda)\right|_{t=0}}{(n+1)!} .
\en
It is helpful to first evaluate
\eq
[\pp_t + \ap[_t], \dpo[_t] \lambda] = \dpo[_t] [(\ap[_t]+\partial_t)(\lambda)].
\en
(Note that both $\dpo[_t] \lambda$ and $\dpo[_t] \ap[_t](\lambda)$ are Hamiltonian
vector fields.)\\[5pt]
\emph{Proof}: (we suppress the $t$-subscripts and the explicit $t$-dependence
of $\lambda$)
\eqa
[\pp_t + \ap, \dpo \lambda](f)
& = & \pp_t( \{ f , \lambda \} ) + \ap( \{ f , \lambda \} ) - \{ \ap(f) ,
      \lambda \}  \nn
& = & \fp(f,\lambda) - (\dpo\ap)(f,\lambda) + \{ f , \ap(\lambda) \}\nn
& = & \dpo (\ap(\lambda))(f).
\ena
We have repeatedly used the definition of $\dpo$
(\ref{hamvec}) and, in the last step, (\ref{fs}).
Now we can use (\ref{useful}) to evaluate
\eqa
\rho^*_{a+d\lambda} - \rho^*_a
 & = & \left.\left(e^{\pp_t + \ap[_t] + \dpo[_t] \lambda} - e^{\pp_t + \ap[_t]}\right)
 e^{-\pp_t}\right|_{t=0}\nn
 & = &
 \sum_{n=0}^\infty \left.\frac{\dpo[_t](\ap[_t] + \pp_t)^n(\lambda)}{(n+1)!} 
 e^{\pp_t + \ap[_t]}e^{-\pp_t}\right|_{t=0} + o(\lambda^2). \quad\Box
\ena
\paragraph{Quantum:}
We would like to proof (\ref{Dadl}) and (\ref{tildehatlambda})
\eq
\DD_{a+d\lambda} - \DD_a= (\frac{1}{i\hbar} \ds\Lambda)\circ\DD_a
+ o(\lambda^2),
\qquad
\Lambda(\lambda,a) =
\sum_{n=0}^\infty \frac{(\as[_t] + \pp_t)^n(\hat\lambda)|_{t=0}}{(n+1)!}.
\label{toshow}
\en
First we evaluate
\eq
[\pp_t + \as[_t], \ds[_t] \hat\lambda] = \ds[_t] [(\as[_t] +
\partial_t)(\hat\lambda)].
\en
(Note that both $\ds[_t] \hat\lambda$ and $\ds[_t] \as[_t](\hat\lambda)$
are inner derivations. Also note that the proof of this equation is in principle much harder than
its semi-classical counterpart, since we are now dealing with differential
operators of arbitrary degree.)
\\[5pt]
\emph{Proof}: (we suppress the $t$-subscripts and the explicit $t$-dependence
of $\lambda$)
\eqa
[\pp_t + \as, \ds \hat\lambda](f) & = & \pp_t[f,\hat\lambda]_\star +
 	\as (\ds(\hat\lambda) f)
	- \ds(\hat\lambda)(\as(f)) \nn
	& = & \fs(f,\hat\lambda) - \fs(\hat\lambda,f) +
	\as ([f,\hat\lambda]_{\star})
	-[\as (f),\hat\lambda]_{\star} \nn
	& = & \fs(f,\hat\lambda) - \fs(\hat\lambda,f)
	-(\ds\as)(f,\hat\lambda) + [f,\as \hat\lambda]_\star
	+ (\ds\as)(\hat\lambda,f) \nn
	& = & \ds(\as(\hat\lambda))(f).
\ena
The desired result (\ref{toshow}) follows now from (\ref{useful}) as in
the semi-classical case.\quad$\Box$

\end{document}